\documentclass[article, onecolumn, 11pt]{IEEEtran}

\usepackage[english]{babel}
\usepackage[utf8]{inputenc}
\usepackage{amsmath}
\usepackage{graphicx}
\usepackage[colorinlistoftodos]{todonotes}
\usepackage{hyperref}

\usepackage{multirow}

\usepackage{bm}

\usepackage{graphicx}
\usepackage{color}
\usepackage{placeins}
\usepackage{float}
\usepackage{tabularx,colortbl}

\usepackage{amsfonts}
\usepackage{subfigure}
\usepackage{bm}
\usepackage{url}
\usepackage{multirow}

\usepackage{amssymb}
\usepackage{moresize}
\usepackage{caption}

\usepackage{algorithm}
\usepackage[noend]{algpseudocode}
\makeatletter
\def\BState{\State\hskip-\ALG@thistlm}
\makeatother
\usepackage{amssymb}
\newtheorem{thProth}{Theorem}
\newtheorem{defin}{Definition}

\newtheorem{prop}{Proposition}

\usepackage{cite}

\begin{document}

\title{Multivariate Cryptosystems for Secure Processing of Multidimensional Signals\thanks{This work is partially funded by the Agencia Estatal de Investigación (Spain) and the European Regional Development Fund (ERDF) under projects WINTER (TEC2016-76409-C2-2-R), by the Xunta de Galicia and the European Union (European Regional Development Fund - ERDF) under projects Agrupación Estratéxica Consolidada de Galicia accreditation 2016-2019 and Red Temática RedTEIC 2017-2018, and by the EU H2020 Programme under project WITDOM (project no. 644371).}}

\author{Alberto~Pedrouzo-Ulloa,~\IEEEmembership{Student Member,~IEEE,}
  Juan~Ram\'on~Troncoso-Pastoriza,~\IEEEmembership{Member,~IEEE,}
  and~Fernando~P\'erez-Gonz\'alez,~\IEEEmembership{Fellow,~IEEE}\thanks{A. Pedrouzo-Ulloa and F. P\'erez-Gonz\'alez are with the Department of Signal Theory and Communications of the University of Vigo, Vigo, 36310 Spain e-mail:\{apedrouzo,fperez\}@gts.uvigo.es.
J.R. Troncoso-Pastoriza is with the Laboratory for Communications and Applications 1 at the \'{E}cole Polytechnique F\'{e}d\'{e}rale de Lausanne, Lausanne, Switzerland e-mail: juan.troncoso-pastoriza@epfl.ch.}}

\maketitle

\begin{abstract}
Multidimensional signals like 2-D and 3-D images or videos are inherently sensitive signals which require privacy-preserving solutions when processed in untrustworthy environments, but their efficient encrypted processing is particularly challenging due to their structure, dimensionality and size. This work introduces a new cryptographic hard problem denoted m-RLWE (multivariate Ring Learning with Errors) which generalizes RLWE, and proposes several relinearization-based techniques to efficiently convert signals with different structures and dimensionalities. The proposed hard problem and the developed techniques give support to lattice cryptosystems that enable encrypted processing of multidimensional signals and efficient conversion between different structures. We show an example cryptosystem and prove that it outperforms its RLWE counterpart in terms of security against basis-reduction attacks, efficiency and cipher expansion for encrypted image processing, and we exemplify some of the proposed transformation techniques in critical and ubiquitous block-based processing applications.\end{abstract}

\begin{IEEEkeywords}
Secure Signal Processing, Lattice Cryptography, Homomorphic Cryptography, Multidimensional Signal Processing, Unattended Secure Processing.
\end{IEEEkeywords}

\IEEEpeerreviewmaketitle

\section{Introduction}
\label{sec:intro}
\IEEEPARstart{I}{n} recent years, we have witnessed an increasing interest in the research of schemes enabling operations with encrypted data. All these solutions are based on Secure Computation techniques, which aim at achieving privacy-preserving solutions for secure processing of sensitive signals \cite{LEB13}. Most of these approaches are based on homomorphic encryption, and rely on Paillier cryptosystem~\cite{Pa99} as the basic block for performing encrypted additions between ciphertexts and multiplications between a ciphertext and a cleartext. This approach can mainly cope with encrypted linear transforms~\cite{BPB09a} with known (cleartext) coefficients.

Gentry's seminal work \cite{Gentry09} introduces a new family of cryptosystems enabling FHE (Fully Homomorphic Encryption) schemes that can perform both additions and multiplications in the encrypted domain, while being resilient against quantum cryptanalysis. Despite the relevance of their theoretical contribution, current FHE schemes are not entirely practical for real scenarios~\cite{Gama16}, so the most promising alternative relies on SHE (Somewhat Homomorphic Encryption) schemes, which have been shown \cite{TGP13} to be able to efficiently work with encrypted signals and encrypted transform coefficients. SHE operations are not limited to binary circuits but can be extended to arithmetic circuits over $\mathbb{Z}_t$, with $t\geq2$, they are more efficient and less expansive. As a counterpart, while FHE schemes can perform an unbounded number of encrypted operations (as binary circuits), SHE schemes can cope only with a limited number of consecutive encrypted operations over the same ciphertext; nevertheless, in most real scenarios, the maximum number of operations that have to be performed on the encrypted data can be previously known, so SHE naturally fits.

The main drawback of FHE and SHE that drives most of the current research efforts in the field of Secure Signal Processing~\cite{LEB13} is their large cipher expansion. Several recent proposals are aimed at mitigating this effect, by introducing packing and unpacking steps that allow to encrypt several messages in only one ciphertext \cite{TKCL07}; hence, cipher expansion can be lowered, at the cost of increasing the computational complexity of the different cryptographic primitives. Nevertheless, among the signal processing applications, those working with images or higher dimensional signals are much more demanding in this sense, as the computational cost and cipher expansion of typical SHE cryptosystems becomes unaffordable for them.

In order to address this problem, in this work we propose and analyze a new hard problem (initially introduced in the conference paper \cite{PTP2015}), denoted $m$-RLWE (multivariate Ring Learning with Errors) that better suits multidimensional signals (e.g., 2-D and 3-D images, video, ...). By rooting the used SHE cryptosystems in this hard problem, we show that we can achieve a reduction of both the computational cost and the cipher expansion along with an increase in the security when working with multidimensional signals. This is so due to the more compact and efficient representation of the signals that outperforms the direct use of packing and unpacking steps in RLWE-based cryptosystems. Furthermore, we show that the use of $m$-RLWE is compatible with other methods, so it can be combined with the previously mentioned packing techniques and CRT (Chinese Remainder Theorem) \cite{DPS96}, which can be leveraged for parallelizing cleartext operations under encryption~\cite{BV11aJ, SV14, NTT2015}. We therefore achieve our \emph{first goal of efficient and practical encrypted processing of multidimensional signals}.

Besides its benefits for multidimensional signals, it must be noted that the $m$-RLWE problem yields further degrees of freedom which can be leveraged to exploit additional structures (not necessarily related to the dimensions of the data) in the data or operations. These structures can be recognized, for example, when processing several signals in parallel or when applying block-wise operations. Therefore, we can achieve \emph{performance and security gains with respect to RLWE in a variety of applications}, especially comprising multi-scale approaches~\cite{PTP16, MMOP07}; these are used, among others, in disciplines like geology, astrophysics, biology, imagery, medicine; being the latter one of the most relevant due to its privacy constraints. Furthermore, $m$-RLWE also \emph{enables a new type of homomorphic operations which are independent of the dimensions presented on both the signals and scenarios}.

\subsection{Post-Quantum Cryptography}

As we have highlighted, Somewhat and Fully Homomorphic Cryptosystems appear as a promising solution enabling both encrypted additions and multiplications, but this is not their only advantage. As a byproduct of being based on hard problems over lattices, they can be proven secure against classical and quantum computers~\cite{Bernstein08}.

Since the introduction of Shor's algorithm~\cite{Shor94}, it is known that some problems which were considered secure against classical adversaries can be efficiently solved by means of a quantum computer~\cite{Bernstein08}. Among these problems, we can mention integer factorization and (elliptic-curve) discrete logarithm, which are the basis of the current most widespread cryptosystems (RSA, Paillier or El Gamal).
Lattice-based cryptography yields the most suited solution to achieving both resilience against quantum attacks and, at the same time, operate on encrypted information.

The quantum-resistance property is another driver for our goal of providing more efficient schemes which can deal with real problems and, additionally, can stand as future-proof against quantum computers.

\subsection{Main Contributions}
Here we summarize and briefly describe our contributions:

\begin{itemize}
\item We propose and study a hard problem called $m$-RLWE (which we introduced in~\cite{PTP2015}) (see Section~\ref{sec:mrlwe}). We give further insights on the structure and features of this problem, relating it with its RLWE counterpart and exemplifying how a homomorphic cryptosystem can be based on this assumption (see Section~\ref{sec:applic}).

\item We present a toolset of multidimensional quantum-resistant secure operations enabled by the $m$-RLWE problem (see Section~\ref{sec:applic}), comprising: a) better encrypted packing of information, b) unattended encrypted divisions without resorting to interactive protocols, and c) multi-scale approaches as wavelet transforms and pyramids.

\item We analyze the use of pre- and post-processing to enable unattended packed and block-processing operations. Additionally, NTTs (Number Theoretic Transforms) are proposed as a means to optimize the encrypted operations (see Section~\ref{sec:scheme}).

\item We develop strategies to homorphically modify the structure of ciphertexts by incorporating some additional information, and without the need of an interactive protocol with the secret key owner, hence enabling different types of unattended secure operations (see Section~\ref{sec:switchingkey}).

\item We evaluate and compare our scheme with previous solutions for several encrypted image processing applications (see Section~\ref{sec:perform}).
\end{itemize}

\subsection{Notation and structure}
\label{sec:notation}

We represent vectors and matrices by boldface lowercase and uppercase letters, respectively. Polynomials are denoted with regular lowercase letters, omitting the polynomial variable (e.g., $a$ instead of $a(x)$) whenever there is no ambiguity.
We indicate the variable of polynomial rings to avoid confusion between univariate and multivariate rings, following a recursive definition of multivariate modular rings: $R_q[x] = \mathbb{Z}_q[x]/(f(x))$ denotes the polynomial ring in the variable $x$ modulo $f(x)$ with coefficients belonging to $\mathbb{Z}_q$. Analogously, $R_q[x, y] = (R_q[x])[y]/(f'(y))$ is the bivariate polynomial ring with coefficients belonging to $\mathbb{Z}_q$ reduced modulo $f(x)$ and $f'(y)$. In general, $R_q[x_1, \dots, x_m]$ (resp. $R[x_1, \ldots, x_m]$) represents the multivariate polynomial ring with coefficients in $\mathbb{Z}_q$ (resp. $\mathbb{Z}$) and the $m$ modular functions $f_i(x_i)$ with $1 \leq i \leq m$. The polynomial $a$ can also be denoted by a column vector $\bm{a}$ whose components are formed by the corresponding polynomial coefficients.
Finally, $\bm{a} \cdot \bm{s}$ is the scalar product between the vectors $\bm{a}$ and $\bm{s}$, whose components can belong to the integers or to a polynomial ring.

The rest of the paper is structured as follows: Section~\ref{sec:prelim} revisits some basic concepts of homomorphic cryptosystems and the underlying hard problems, together with an adapted definition of the multivariate RLWE problem; Section~\ref{sec:mrlwe} discusses the hard problems on which the multivariate RLWE problem bases its security. We also give some insights on the relation between RLWE and $m$-RLWE formulations. Section~\ref{sec:applic} introduces a set of possible encrypted unattended applications for which $m$-RLWE brings about notable optimizations; Section~\ref{sec:scheme} includes the description of the main tools proposed in this work. Section~\ref{sec:switchingkey} proposes an optimization which enables to homomorphically update the structure of the ciphertexts, and Section~\ref{sec:perform} compares the performance and security of our methods with respect to solutions based on RLWE and Paillier.

\section{Preliminaries}
\label{sec:prelim}
The state of the art in FHE is based on the Learning with Errors (LWE)~\cite{BLPRS13} and Ring Learning with Errors (RLWE) problems~\cite{LPR10J}, which have proven security reductions from hard lattice problems. Both RLWE leveled cryptosystems~\cite{BV11aJ}, which enable the homomorphic execution of a bounded-degree polynomial function, and scale-invariant leveled cryptosystems based on RLWE produce the currently most efficient FHE systems~\cite{Brakerski12, FV12, BLLN13}.

Both RLWE and LWE have a similar formulation, that Brakerski~\emph{et al.} generalized to a common General Learning with Errors (GLWE) problem \cite{BV11aJ}. We recall a slightly adapted informal definition of GLWE, as the basis for our schemes introduced in the next sections:

\begin {defin}[GLWE problem~\cite{BV11aJ}]
Given a security parameter $\lambda$, an integer dimension $l=l(\lambda)$, two univariate polynomial rings $R[x]=\mathbb{Z}[x]/(f(x))$, $R_q[x]=\mathbb{Z}_q[x]/(f(x))$ with $f(x)=x^n+1$, $q=q(\lambda)$ a prime integer, $n=n(\lambda)$ a power of two, and an error distribution $\chi[x]\in R_q[x]$ that generates small-norm random univariate polynomials in $R_q[x]$, GLWE$_{l,f,q,\chi}$ relies upon the computational indistinguishability between pairs of samples $( \bm{a}_i, b_i = \bm{a}_i \cdot \bm{s} + t\cdot e_i )$ and $( \bm{a}_i, u_i )$, where $\bm{a}_i\leftarrow R^l_q[x]$, $u_i\leftarrow R_q[x]$ are chosen uniformly at random, $\bm{s}\leftarrow \chi^l[x]$ and $e_i \leftarrow \chi[x]$ are drawn from the error distribution, and $t$ is an integer relatively prime to $q$.
\end {defin}

When $n=1$, GLWE becomes the standard LWE$_{l,q,\chi}$, and when $l=1$ it reduces to RLWE$_{q,f,\chi}$. LWE-based cryptosystems yield huge expansion factors and are computationally demanding, reason why RLWE was defined as an algebraic version of LWE, trading subspace dimensionality for polynomial ring order (using an ideal ring), and achieving huge efficiency improvements. As for the generic GLWE ($n>1$ and $l>1$), Brakerski~\emph{et al.}~\cite{BV11aJ} speculate that it is hard for $n\cdot l=\Omega\left(\lambda\log (q/B)\right)$, where $B$ is a bound on the length of the elements output by $\chi[x]$. It must be noted that despite the efficiency improvement, there are no known attacks in RLWE that get a substantial advantage with respect to attacks to LWE.\footnote{For a formal definition of the GLWE problem and proofs of security reductions for RLWE and LWE, we refer the reader to~\cite{LPR10J,BV11aJ,BLPRS13}.} Hence, the currently most efficient homomorphic cryptosystems are based on RLWE, particularly BGV~\cite{BV11aJ,BV11b} and NTRU~\cite{LTV13}, together with their scale-invariant counterparts FV~\cite{FV12} and YASHE~\cite{BLLN13}; depending on the requirements of the specific application, the optimal choice of the used RLWE-based cryptosystem can be different as analyzed by Costache and Smart in \cite{CS16}.

We now introduce our extension of RLWE to the multivariate case, and build a set of tools to enable efficient and unattended multidimensional encrypted processing.

\section{Multivariate Ring Learning with Errors}
\label{sec:mrlwe}
In this section, we recall the definition of $m$-RLWE and sketch the proof for the hardness of the multivariate Ring Learning with Errors problem. In \cite{PTP2015, PTP16mrlwe} we proposed a generalization of RLWE as a new problem called $m$-RLWE (multivariate Ring Learning with Errors), providing an exemplary new cryptosystem based on it, especially designed for encrypted image filtering. The $m$-RLWE hardness assumption is especially useful for working with multidimensional signals; for simplicity of the exposition, we present cryptosystems extending Lauter's cryptosystem \cite{LNV11} (a simpler non-leveled version of BGV), but the same methodology can be applied to any other RLWE-based cryptosystem as those previously cited. The formulation of the $m$-RLWE problem is the following:

\begin {defin}[Multivariate RLWE ($m$-RLWE) problem~\cite{PTP2015, PTP16mrlwe}] \label{def:mrlwe}
Given a multivariate polynomial ring $R_q[x_1, \ldots, x_m]$ with $f_j(x)=x_j^{n_j} + 1$ for $j = 1, \ldots, m$ and an error distribution $\chi[x_1, \ldots, x_m]\in R_q[x_1, \ldots, x_m]$ that generates small-norm random multivariate polynomials in $R_q[x_1, \ldots, x_m]$, $m$-RLWE relies upon the computational indistinguishability between samples $(a_i, b_i = a_i \cdot s + t\cdot e_i)$ and $(a_i , u_i)$, where $a_i$, $u_i$ $\leftarrow R_q[x_1, \ldots, x_m]$ are chosen uniformly at random from the ring $R_q[x_1, \ldots, x_m]$; $s,e_i \leftarrow \chi[x_1, \ldots, x_m]$ are drawn from the error distribution, and $t$ is relatively prime to $q$.
\end {defin}

We can state a worst-case to average-case reduction from the shortest vector problem (SVP) over ideal lattices to the $m$-RLWE problem. For convenience, we particularize the reduction presented in~\cite{PTP16mrlwe} to the more specific $m$-RLWE definition included in this manuscript.

Let $\lambda$ be the security parameter and $n = 2^{\lfloor \log{\lambda} \rceil - 1}$ with $\lambda \in \mathbb{N}$. Consider also an upper bound $B$ with overwhelming probability over the length (in Euclidean norm) of the elements sampled from $\chi[x_1, \ldots, x_m]$ (see Lemma $1$ from~\cite{BV11b}). We can state the following theorem:
\begin{thProth}[Theorem 1 from~\cite{PTP16mrlwe} particularized to Definition~\ref{def:mrlwe}]
Let the rings $R[x_1, \ldots, x_m]$ (defined over $\mathbb{Z}$) and $R_q[x_1, \ldots, x_m]$ already introduced in Definition~\ref{def:mrlwe}. Let $q \equiv 1 \bmod{ 2 \cdot \max{\{n_1, \ldots, n_m\}}}$ be a poly$(n)$-bounded prime. Consider also the error distribution $\chi[x_1, \ldots, x_m]$ with parameter $r \geq \omega(\sqrt{\log{n}})$ and whose output elements belong to $R$ and are of length at most $B$ with overwhelming probability. Then, there exists a polynomial-time quantum reduction from $\tilde{\mathcal{O}}(nq/B \cdot {(n(l+1)(\log{n(l+1)}))}^{1/4})$-approximate SVP over ideal lattices in $R$ to $m$-RLWE (Definition~\ref{def:mrlwe}), given only $l$ samples $(a_i, b_i)$, which runs in time poly($n$, $q$, $l$).
\end{thProth}

The proof for the security reduction of $m$-RLWE can be divided in two fundamental blocks:
\begin{itemize}
\item \emph{Hardness Search-LWE}, which describes a quantum reduction from approximate SVP (Shortest Vector Problem) on ideal lattices over the tensor product of an arbitrary number of ring of integers to the search version of $m$-RLWE. The search version tries to recover the secret key $s$.
\item \emph{Pseudorandomness of $m$-RLWE}, which describes a reduction from the search version to the decision variant of the problem (which is also more amenable for cryptographic applications). A simplified version of the decision version of the $m$-RLWE problem is included in Definition~\ref{def:mrlwe} (for clarity on the exposition, it only considers polynomial rings whose modular functions have the form $f(x) = x^n + 1$, with $n$ a power of two).
\end{itemize}

The full proof can be found in our pre-print~\cite{PTP16mrlwe}. In this work, we want to give some further insights about the analogous structure between an $m$-RLWE sample and an RLWE sample. We also discuss the security relation between both problems in terms of the indistinguishability assumption. For this purpose, we revisit and expand the propositions introduced in \cite{PTP2015}, concerning the distributions of both RLWE and the presented $m$-RLWE. We do this by first analyzing the distribution of the bivariate RLWE problem, that we later generalize by induction to $m$-variate polynomial rings.

\subsection {Bivariate RLWE ($2$-RLWE)}
\label{sec:2RLWE}
The bivariate version of RLWE can be described by substituting the polynomial ring by a bivariate one $R_q[x,y] = (R_q[x])[y]/(f'(y))$, such that the error distribution $\chi[x,y]$ generates also low-norm bivariate polynomials from $R_q[x,y]$:

\begin {defin}[Bivariate RLWE ($2$-RLWE)~\cite{PTP2015, PTP16}]
Given a bivariate polynomial ring $R_q[x,y]$ with $f(x)=x^{n_1}+1,f'(y)=y^{n_2}+1$ and an error distribution $\chi[x,y]\in R_q[x,y]$ that generates small-norm random bivariate polynomials in $R_q[x,y]$, $2$-RLWE relies upon the computational indistinguishability between samples $(a_i, b_i = a_i \cdot s + t\cdot e_i)$ and $(a_i , u_i)$, where $a_i$, $u_i$ $\leftarrow R_q[x, y]$ are chosen uniformly at random from the ring $R_q[x,y]$, $s,e_i \leftarrow \chi[x,y]$ are drawn from the error distribution, and $t$ is relatively prime to $q$.
\end {defin}

Informally, $2$-RLWE is to GLWE \cite{BV11aJ} what RLWE is to LWE, as we are trading (for a second time) subspace dimensionality for a higher polynomial ring degree, therefore increasing the security of regular RLWE and improving on performance with respect to GLWE. 

The dimensionality of the noise distribution is now $n=n_1 \cdot n_2$, and we preserve most of the relevant properties of the used ideals by considering the bivariate rings as the tensor product (as $R$-modules) of the ring of integers of a cyclotomic field. Additionally, it can be seen that for the coefficient embedding the ideal lattices equivalent to this product ring are generated by block negacyclic matrices of dimension $n=n_1\cdot n_2$. We now enunciate the following proposition about the distribution of a $2$-RLWE sample.

\begin{prop}[Prop.~1 in~\cite{PTP2015}]
A $2$-RLWE sample with $n_x = n$ and $n_y = l$ is indistinguishable from a sample belonging to RLWE with $n_z = l\cdot n$.
\label{prop:equivalence2RLWE}
\end{prop}
For the proof of Prop.~\ref{prop:equivalence2RLWE} we separately analyze the distribution of both a RLWE sample and a $2$-RLWE sample. In order to do this, we use the polyphase decomposition of the involved signals, with the particularity that due to the cryptosystem requirements, which assume polynomials modulo $1 + z^n$, we must work with negacyclic convolutions~\cite{Dav12}. Next, the details of the proof of Prop.~\ref{prop:equivalence2RLWE} are included.

\subsubsection{RLWE sample}
Let us consider a typical RLWE sample $(a, b = a \cdot s + e)$, where $a, b \in R_q[z]$ with $f(z) = z^{ln} + 1$ and $e \leftarrow \chi[z]$. We can write the polynomial $b(z) = \sum_{k = 0}^{l-1} z^k b_k(z^l)$ as its decomposition according to its first $l$ polyphase components $b_k(z) = \sum_{m = 0}^{n-1} b[lm + k] z^{m}$ with $k = 0, 1, \ldots, l-1$, where

\begin{equation} 
b_k(z) = e_k(z) + \sum_{i + j = k}a_i(z)s_j(z) + z\sum_{i + j = l + k}a_i(z)s_j(z),
\label{eq:polyRLWE}
\end{equation}
and $e_k(z) = \sum_{m = 0}^{n - 1} e[lm + k] z^m$, $a_i(z) = \sum_{m = 0}^{n - 1} a[lm + i]$ and $s_j(z) = \sum_{m = 0}^{n - 1} s[lm + j]$ where $k, i, j = 0, \ldots, l - 1$ are, respectively, the $k$-th, $i$-th and $j$-th polyphase components of the polyphase decomposition in $l$ components of $e(z)$, $a(z)$ and $s(z)$, respectively.

\begin{figure}[h]
\centering
\includegraphics[width=6in]{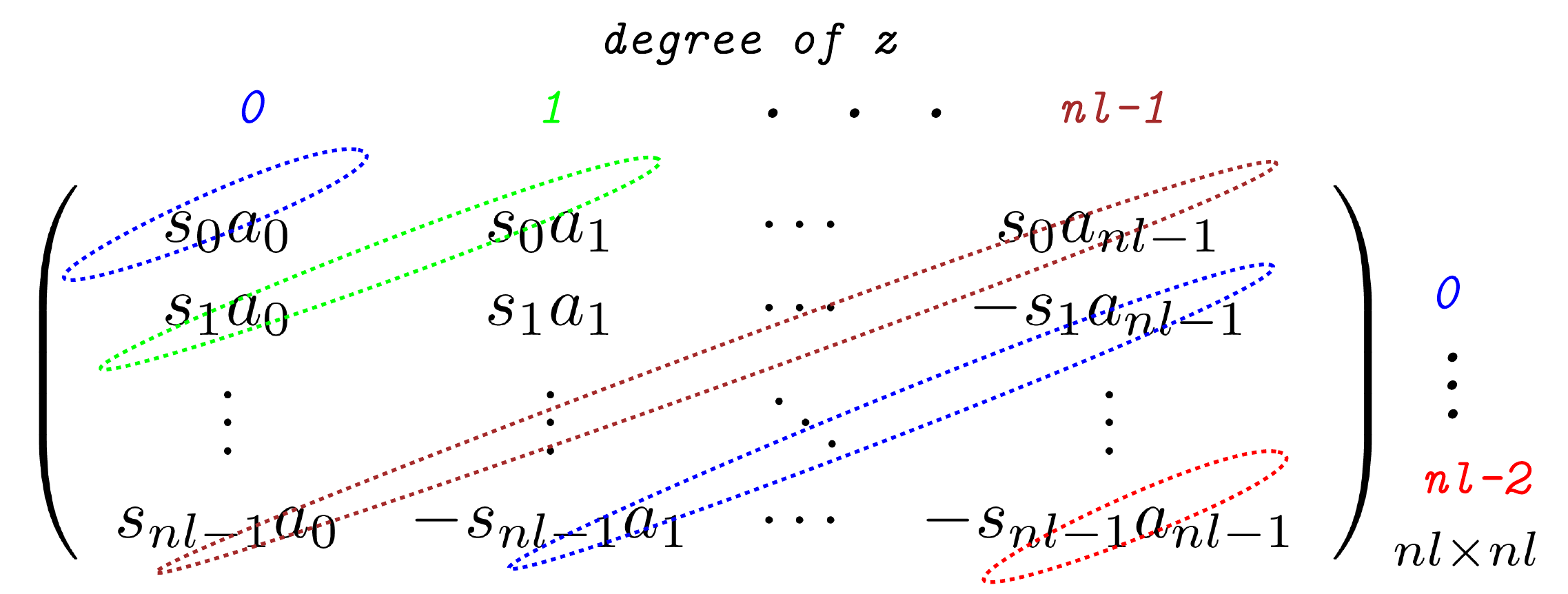} 
\caption{Product combinations for the coefficients of a RLWE sample.}
\label{fig:matrizRLWE}
\end{figure}

Hence, each RLWE sample can be represented as a set of $l$ equations with $(n-1)$-degree polynomials. As the convolutions are negacyclic, each one of the $nl$ coefficients of the RLWE sample is equal to the summation of $nl$ different products of coefficients from $a(z)$ and $s(z)$, plus a noise sample from $e(z)$. In those sums of products, the combination of the different coefficients $a_i s_j$ present for each coefficient of $b(z)$ is unique, so we have $n^2l^2$ different combinations of products for all equations. Figure~\ref{fig:matrizRLWE} graphically shows, in matrix form, the product combinations for each polynomial coefficient.

\subsubsection{$2$-RLWE sample} 
Consider now a $2$-RLWE sample $(a, b = a \cdot s + e)$ with $a, s \leftarrow R_q[x, y]$, $e \leftarrow \chi[x, y]$, $f_x(x) = x^n + 1$ and $f_y(y) = y^l + 1$.

If we denote the coefficients of $y^k$ for each signal with $s_k(x)$, $b_k(x)$, $e_k(x)$, $s_k(x)$, respectively, we have the following expression for $0 \leq k < l$:

\begin{equation} 
b_k(x) = e_k(x) + \sum_{i + j = k} a_{i}(x) s_j(x) - \sum_{i + j = n + k} a_{i}(x) s_j(x).
\label{eq:poly2RLWE}
\end{equation}

That is, we can see it as a polyphase decomposition in which the coefficients are shuffled in blocks and subtracted prior to the extraction of each phase.

\begin{figure}[h]
\centering
\includegraphics[width=6in]{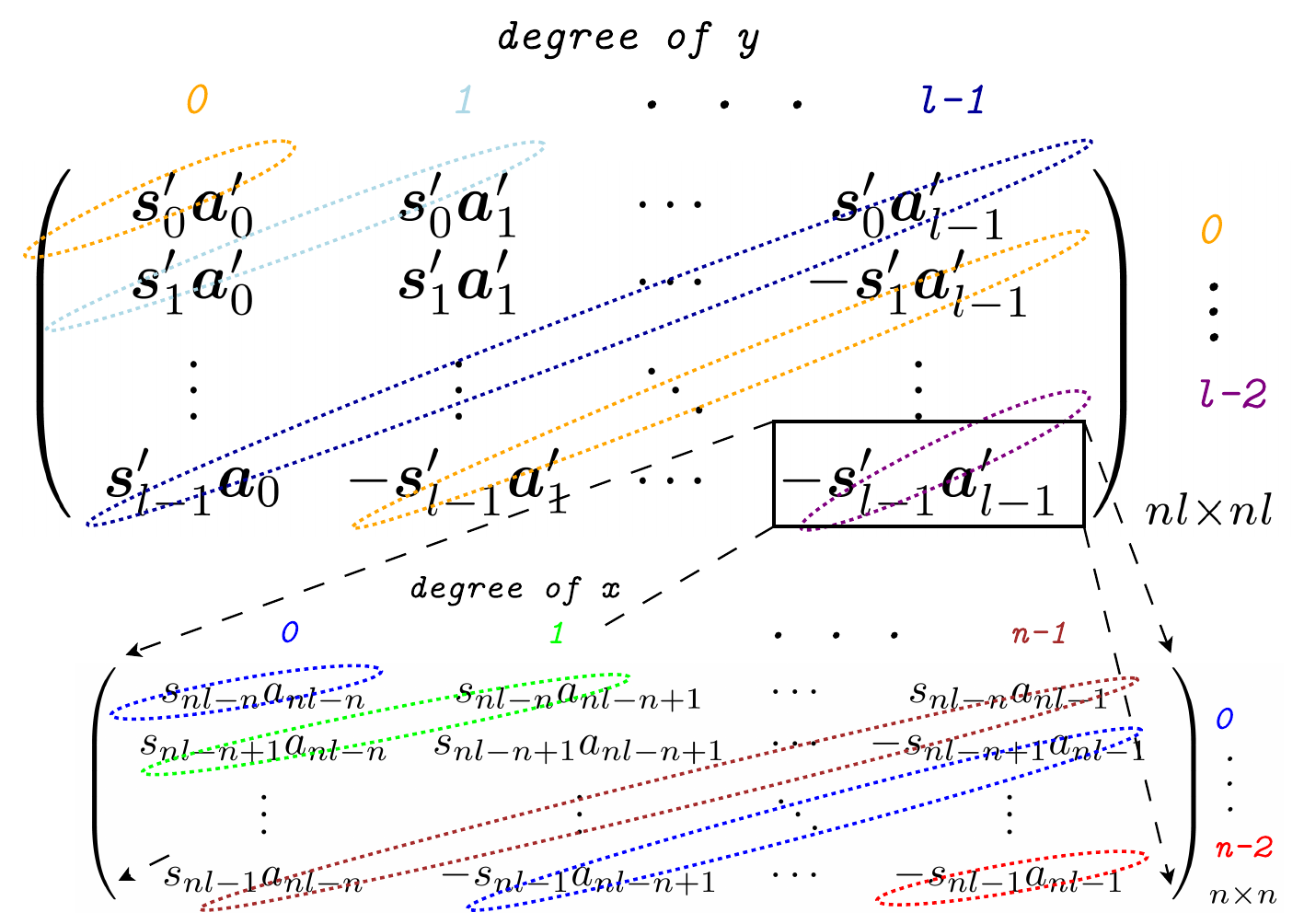} 
\caption{Product combinations for the coefficients of a $2$-RLWE sample.}
\label{fig:matriz2RLWE}
\end{figure}

At this point, we can see the parallelism between Eqs.~\eqref{eq:polyRLWE} and~\eqref{eq:poly2RLWE}. In order to show that they are fully equivalent expressions in terms of the sample distribution, let us build the $2$-RLWE vectors $\bm{a}$ and $\bm{s}$ as the following block composition of the $a_i$ and $s_j$ coefficients of the RLWE sample vectors:

\[
\bm{a} = \left(a_0, a_1, \ldots,a_{nl-1}\right)_{1\times nl} = \left(\bm{a}'_0,\bm{a}'_1, \ldots, \bm{a}'_{l-1}\right)_{1\times nl}, \]
\[
\bm{s}^T = \left(s_0, s_1, \ldots, s_{nl-1}\right)_{1\times nl} = \left({\bm{s}'_0}^T,{\bm{s}'_1}^T,\ldots,{\bm{s}'_{l-1}}^T\right)_{1\times nl},\]
where the involved $\bm{a}'_i$ and $\bm{s}'_i$ are respectively row and column vectors of length $n$. Using these vectors, Figure~\ref{fig:matriz2RLWE} depicts their product combinations in block matrix form, for the $2$-RLWE sample.

We can interpret Eqs.~\eqref{eq:polyRLWE} and \eqref{eq:poly2RLWE} as equivalent ways of expressing the RLWE and $2$-RLWE distributions, respectively; the only difference between both lies in the coefficient ordering of the used $\bm{s}$, $\bm{e}$ and $\bm{a}$.
The coefficients of the $2$-RLWE sample correspond to the summation of the different products of the coefficients of $a$ and $s$, plus a noise sample. As the signal blocks $a_{i}(x)$ and $s_{i}(x)$ do not share any sample with the other blocks $a_{j}(x)$ and $s_{j}(x)$ for $j \neq i$, and all the negacyclic convolutions are performed between different blocks, we can see that all the product combinations are different. Thus, Eqs~\eqref{eq:polyRLWE} and \eqref{eq:poly2RLWE} are perfectly analogous up to coefficient ordering and sign; as they have the same number of equations, both expressions are formed by the summation of different coefficient products of $a$ and $s$, and finally, they have $n^2l^2$ different combinations of products in total. This is graphically shown in Figures~\ref{fig:matrizRLWE} and~\ref{fig:matriz2RLWE}.

Furthermore, as $\bm{s}$ and $\bm{e}$ have a symmetrical distribution and $\bm{a}$ is uniformly chosen, the distribution of both samples is exactly the same.

\subsection{Multivariate RLWE ($m$-RLWE)}
\label{sec:mRLWE}
Resorting to the recursive definition of multivariate polynomial rings (see Section~\ref{sec:notation}), the Bivariate RLWE problem can be seamlessly extended to multivariate polynomials ($m$-RLWE) with $m>2$, by recursively applying the proposed modification to the general GLWE problem. The formulation is perfectly analogous to the 2-RLWE with rings $R[x_1,\ldots,x_m]$ and $R_q[x_1,\ldots,x_m]$, and error distribution $\chi[x_1,\ldots,x_m]$ (see Definition~\ref{def:mrlwe}).

\begin{prop}[Prop.~2 in~\cite{PTP2015}]
An $m$-RLWE sample with $n_i$ and $f(x_i) = 1 + x_i^{n_i} $ for $ i = 1, \dots, m $ is indistinguishable from a RLWE sample with $n = \prod n_i$.
\label{prop:equivalencemRLWE}
\end{prop}

Whenever the cyclotomic polynomials in each variable $x_i$ have the form $1 + x_i^{n_i}$ (the degree is a power of two), the same procedure sketched above for proving Prop.~\ref{prop:equivalence2RLWE} can be applied to prove the equivalence of one sample from $m$-RLWE (with $n_1, n_2, \ldots, n_m$) and one sample from ($m-1$)-RLWE distributions (with $n_1, n_2, \ldots, n_{m-2}, n_z$), by ``folding'' two variables of the former ($n_{m-1},n_{m}$) onto one variable of the latter ($n_z$). Therefore, Prop.~\ref{prop:equivalencemRLWE} can be proven by induction using the following procedure:

\begin{itemize}

\item First, we have shown the equivalence between one sample RLWE and one sample $2$-RLWE ($n = l_1l_2$).

\item Then, if we assume the equivalence between one sample $(m-1)$-RLWE and one sample RLWE (with $n = n_1n_2 \ldots n_{m-2}n_z$), we have to prove the equivalence between one sample $(m-1)$-RLWE (with $n_1,n_2, \ldots, n_{m-2}, n_z$) and one sample $m$-RLWE (with $n_1,n_2, \ldots, n_{m-2}, n_x, n_y$, where $n_z = n_xn_y$). We only have to account for a recursive application of the previous equations \eqref{eq:polyRLWE} and \eqref{eq:poly2RLWE}. For it, we simply consider that instead of operating with coefficients belonging to the integers, all the involved coefficients are multivariate polynomials with $m-2$ variables and they also have the same modular functions for both the ($m-1$)-RLWE and $m$-RLWE sample. Analogously, for a graphical explanation, we can consider that the elements $a_i$ and $s_j$ in Figures~\ref{fig:matrizRLWE} and~\ref{fig:matriz2RLWE} are also multivariate polynomials with $m-2$ variables or, equivalently, that the matrices for the $a_is_j$ products in $m$-RLWE are block matrices that can be recursively decomposed until reaching RLWE.

\end{itemize}

Thus, through a recursive repetition of the argument for the distribution indistinguishability between one RLWE sample and one $2$-RLWE sample (as stated above), Prop.~\ref{prop:equivalencemRLWE} can be proven.

Regarding the security relation between RLWE and $m$-RLWE in terms of the computational cost for breaking their respective indistinguishability assumptions, we can see that both problems have analogous security reductions from hard problems over ideal lattices, but we cannot assert whether one of them is more computationally difficult than the other or both are equivalent problems. In this sense, the better known attacks for lattice-base reduction do not get noteworthy advantage when comparing ideal lattices and random lattices. Therefore, for our security evaluation (see Section~\ref{sec:perform}), we use the best available attacks for the underlying lattices of each hardness problem (RLWE or $m$-RLWE problem).

Finally, even though the security tradeoffs between RLWE and $m$-RLWE are not conclusive, we want to remark that in the case that both the RLWE and $2$-RLWE problems considered in Proposition~\ref{prop:equivalence2RLWE} were equivalent for more than one sample, the inductive argument provided in this section for $m$-RLWE would apply, and the Proposition~\ref{prop:equivalencemRLWE} would imply that $m$-RLWE and RLWE are also equivalent problems in terms of the computational cost for breaking the indistinguishability assumption.

\section{Applications of $m$-RLWE for Secure Computation}
\label{sec:applic}
This section discusses how the $m$-RLWE problem can enable encrypting multidimensional information while still preserving its structure. As we show, this can be achieved with only a small overhead on cipher expansion with respect to the version in the clear, enabling additive and multiplicative homomorphisms, and reaching higher levels of security when compared to the counterpart protocols using RLWE-based primitives, without reducing the efficiency.

We briefly recall first the example cryptosystem presented in~\cite{PTP2015} and the use of $m$-RLWE for performing encrypted multidimensional linear convolutions. Next, we introduce a set of practical scenarios where the $m$-RLWE problem can produce effective solutions.
These methods are not exclusive for multidimensional signals, so they can also be of benefit to unidimensional signals. Among the proposed solutions, we find a better way to pack the information, we enable encrypted divisions without an interactive protocol, and we implement encrypted versions of several multi-scale algorithms (e.g., wavelet transforms and pyramids) which are widely used in both computer vision and signal processing applications. We provide here a high level description for these solutions, and detail the proposed underlying mechanisms in Section~\ref{sec:scheme}.

\subsection{An example of an $m$-RLWE based Cryptosystem}

Any cryptosystem whose security is based on RLWE (e.g., \cite{BV11aJ, BV11b, LNV11, LTV13, FV12, BLLN13}) could be extended to $m$-RLWE. In \cite{PTP2015}, we extended Lauter \emph{et al.}'s~\cite{LNV11}, due to its efficiency and security, as a basis to exemplify the main properties of a semantically secure $m$-RLWE-based cryptosystem. Table~\ref{tab:table1} summarizes its parameters and primitives. There are currently more efficient choices like FV~\cite{FV12} or BGV~\cite{BV11aJ}, but we prefer to abstract the peculiarities of the high level cryptosystem functions and focus on the actual functionalities that our proposed mechanisms enable.
Our results can be straightforwardly extended to more efficient cryptosystems in case it is required.

\begin{table}[!t] 
\renewcommand{\arraystretch}{1.3}
\caption{Proposed Cryptosystem: Parameters and Primitives}
\label{tab:table1}
\centering \scriptsize

\begin{tabular}{|m{0.45in}|m{0.27in}|m{2.3in}|}
\hline
\multicolumn{3}{|c|}{Parameters}\\
\hline
\multicolumn{3}{|p{3.4in}|}{ 
Let $R_t[x_1, \dots, x_m]$ be the cleartext ring and $R_q[x_1, \dots, x_m]$ as ciphertext's. The noise distribution $\chi[x_1, \dots, x_m]$ in $R_q[x_1, \dots, x_m]$ takes its coefficients from a spherically-symmetric truncated i.i.d Gaussian $\mathcal{N}(\bm{0},\sigma^2\bm{I})$. 
$q$ is a prime $q \equiv 1 \mod{2 \max{\{ n_1, \ldots, n_m \}}}$ (with $n = \prod n_i$), and $t<q$ is relatively prime to $q$.
}\\
\hline

\multicolumn{3}{|c|}{Cryptographic Primitives}\\

\hline
SH.KeyGen& Process & $s, e \leftarrow \chi[x_1,\ldots,x_m]$, $a_1 \leftarrow R_q[x_1, \dots , x_m]$  $sk = s$ and $pk = (a_0 = -(a_1 s + te), a_1)$ \\

\hline
\multirow{2}{*}{SH.Enc}
& Input & $pk = (a_0, a_1)$ and $m \leftarrow R_t[x_1, \dots, x_m]$ \\
\cline{2-3}
& Process & $u, f, g \leftarrow \chi[x_1,\ldots,x_m]$ and the fresh ciphertext is $\bm{c} = (c_0, c_1) = (a_0 u + tg + m, a_1 u + tf)$ \\

\hline
\multirow{2}{*}{SH.Dec}
& Input & $sk$ and $\bm{c} = (c_0, c_1, \dots,c_{\gamma-1})$ \\
\cline{2-3}
& Process & $m = \left( \left(\sum_{i = 0}^{\gamma-1} c_i s^i \right) \mod q \right) \mod{t}$ \\

\hline
\multirow{2}{*}{SH.Add}
& Input & $\bm{c}_0 = (c_0, \dots, c_{\beta-1})$ and $\bm{c}_1 = (c_0', \dots, c_{\gamma-1}')$ \\
\cline{2-3}
& Process & $\bm{c}_{add} = (c_0 + c_0', \dots, c_{\max{(\beta, \gamma)}-1} + c_{\max{(\beta, \gamma)}-1}')$  \\

\hline
\multirow{2}{*}{SH.Mult}
& Input & $\bm{c}_0 = (c_0, \dots, c_{\beta-1})$ and $\bm{c}_1 = (c_0', \dots, c_{\gamma-1}')$ \\
\cline{2-3}
& Process & Using a symbolic variable $v$ their product is $\left( \sum_{i=0}^{\beta-1} c_i v^i \right) \cdot \left( \sum_{i=0}^{\gamma-1} c_i' v^i \right) = \sum_{i=0}^{\beta + \gamma-2} c_i'' v^i$ \\

\hline
\end{tabular}
\end{table}

The cryptosystem in Table~\ref{tab:table1} supports both additions (the smallest ciphertext is previously zero-padded) and multiplications between ciphertexts which are composed by $\gamma\geq2$ ring elements from $R_q[x_1, \dots, x_m]$. This encryption size increases with each multiplication (see Table~\ref{tab:table1}), and it can be brought back to the size of a fresh cipher by means of a relinearization step, which involves using partial encryptions of the secret key (more details can be found in~\cite{BV11aJ,LNV11}, and Section~\ref{sec:switchingkey}).

\paragraph*{\bf{Security and Correctness}}

The security of the cryptosystem is based on the computational difficulty of reducing the $n$-dimensional lattice ($n=\prod n_i$) generated by the secret key, and on the semantic security guaranteed by the underlying $m$-RLWE problem (two encryptions of the same or different plaintexts cannot be distinguished). As for correctness, $q$ must be set such that enough ``space'' is guaranteed to avoid decryption errors produced by wrap-arounds of the performed homomorphic operations.
Due to the analogous (not isomorphic) structure of $m$-RLWE with $n=\prod n_i$ and $n$-degree RLWE (cf. Section~\ref{sec:mRLWE}), bounds for the error norm~\cite{LNV11} are preserved when switching from RLWE to $m$-RLWE, by adjusting the increased dimensionality of the ring elements: for $D$ successive products between fresh ciphertexts and $A$ sums, the needed $q$ for correct decryption is lower-bounded by
\begin{equation}
\label{eq:condq}
q \ge 4(2t \sigma^{2} \sqrt{n_1 n_2 \ldots n_m})^{D+1}(2n_1 n_2 \ldots n_m)^{D/2}\sqrt{A}.
\end{equation}

\paragraph*{\bf{Cryptographic Primitives}}
Although in this work we focus on primitives for homomorphic cryptography, ideal lattices have also been used to develop algorithms for key exchange \cite{Peikert14} and signatures \cite{DDLL13}. Hence, we want to remark that those primitives based on RLWE could be also extended to the $m$-RLWE problem.

\subsection{Encrypted Multidimensional Linear Convolutions}
\label{sec:multi-convolutions}

Unlike RLWE-based cryptosystems, which lack support for multidimensional signals, the proposed cryptosystem~\cite{PTP2015} introduces a natural way to work with multidimensional linear operations. Additionally, it achieves a more compact representation of the data, as it can effectively encrypt one signal value per coefficient of the encryption polynomial. We exemplify here the implementation of different representative encrypted processing operations like convolutions, correlations or filtering, showing the advantages of the proposed cryptosystem compared to its RLWE-based counterpart. Unless otherwise stated, we always consider that all the used signals and filters are encrypted, to fully conceal all the involved elements in an untrustworthy environment.

Convolutions, correlations and filtering can all be expressed as a linear convolution between two $m$-dimensional signals $\bm{X}$ and $\bm{H}$, namely $\bm{Y}[n_1, \dots, n_m] = \bm{X}[n_1, \dots, n_m] \ast \bm{H}[n_1, \dots, n_m]$, which is equivalent to the ring product of the signals represented as multivariate polynomials $y(z_1,\ldots,z_m)=x(z_1,\ldots,z_m) \cdot h(z_1,\ldots,z_m)$. Using the original RLWE-based scheme, an encrypted convolution would comprise encoding each dimension of the two signals separately as elements of the univariate polynomial ring $R_t[z]$, resulting in two $(m - 1)$-dimensional elements $\bm{X}_{n_1, \dots, n_{m - 1}} (z)$ and $\bm{H}_{n_1, \dots, n_{m - 1}} (z)$ of $R^{m-1}_t[z]$. If $N_{n_i, y}$ is the number of samples in dimension $n_i$ for the signal $y$, the number of involved polynomial products is $\prod_{i = 1}^{m - 1} N_{n_i, x} N_{n_i, h}$ (i.e., $N^{2(m - 1)}$ if $N_{n_i,x} = N_{n_i,h} = N$).

Contrarily, with our proposed cryptosystem the convolution can be done through a single polynomial product of the encryptions, homomorphic to the polynomial product of the clear text. In particular, an encrypted image convolution with the proposed cryptosystem would translate into the product of two bivariate polynomial encryptions.

\paragraph*{\bf{Complex signals}}
$m$-RLWE also enables to naturally incorporate one extra variable to cope with complex signals, represented in the polynomial ring $\mathbb{Z}_t[w]/(w^2+1)$, isomorphic to the complex integers ring, where the variable $w$ plays the role of the imaginary unit.

\paragraph*{\bf{Edge Detection Algorithms}}
As an example of multidimensional convolutions, the Sobel operator is frequently used in image processing and computer vision applications as part of edge detection algorithms. Resorting to the homomorphic product property of the $m$-RLWE cryptosystem, we can easily convolve the Sobel kernel (any other different type of kernel could be considered) with the encrypted image (even a 3D image).

Additionally, if the kernel operator is public, it can be in the clear when convolving it with the encrypted image, hence being its homomorphic execution even more efficient.

\subsection{Better Encrypted Packing}

It can be seen that for practical image processing scenarios it is not so common to filter the whole image. In fact, images are usually divided in different blocks and independent operations are applied to each block.

The approach introduced in~\cite{PTP2015} applied to this scenario would encrypt each block separately. However, this would not benefit from the use of $2$-RLWE ($m$-RLWE with bivariate polynomials) because we would not be encrypting the whole image in only one ciphertext.

In order to preserve the same security (related to the dimension of the underlying bivariate lattices) as in \cite{PTP2015}, we propose different mechanisms to pack the information by exploiting the block-structure of the operation and restructuring the signals into ``virtual'' dimensions that can be leveraged by an $m$-RLWE encryption.

\begin{figure}[h]
\centering
\includegraphics[width=6in]{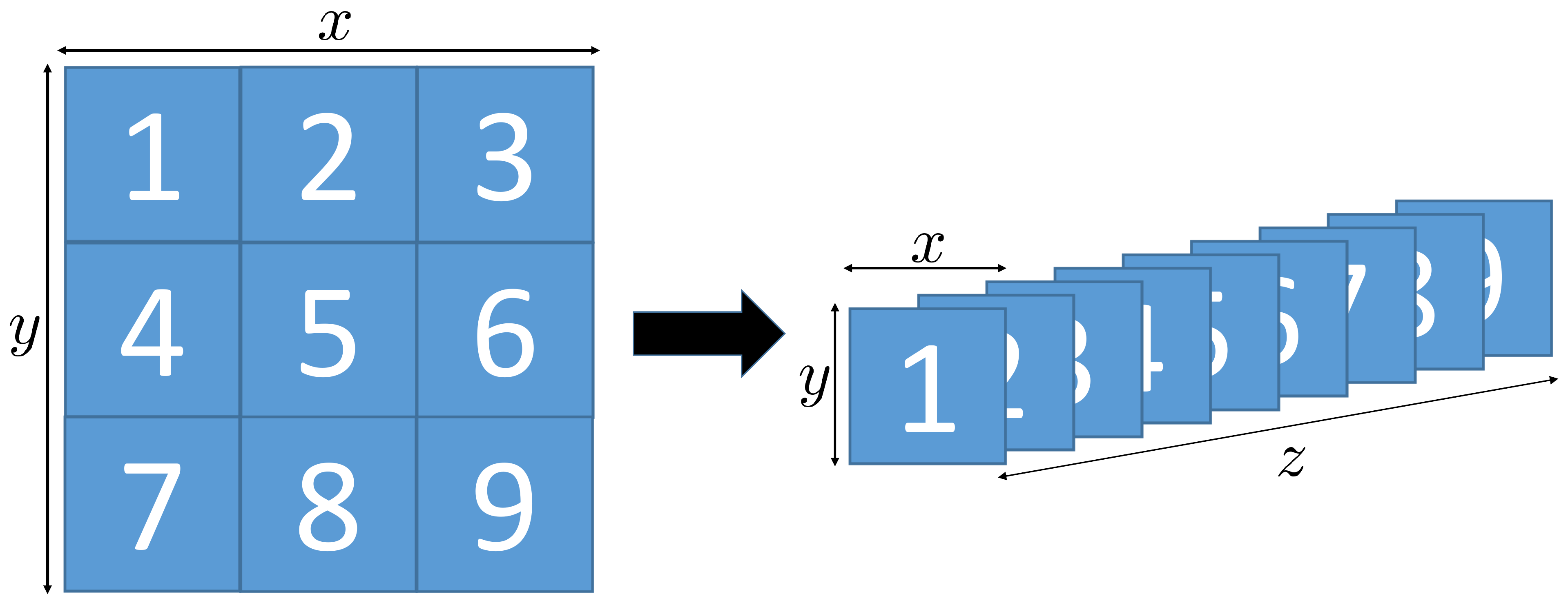} 
\caption{Indexing a set of blocks with $3$-RLWE.}
\label{fig:packing}
\end{figure}

Instead of encrypting each block independently, we can consider one additional polynomial variable for representing the image like a video sequence where each frame corresponds to a different image block (see Figure~\ref{fig:packing}). Therefore, we can get an optimal packing of the information while preserving and exploiting the block structure in the encrypted domain.

The division of an image into blocks is not the unique additional dimension that we can consider and, in any real application, we can also take into account the number of plaintext signals which we want to work with. One would traditionally have to encrypt each signal in a different ciphertext, ending up with as many ciphertexts as plaintext signals in the process. To this end, we can use $m$-RLWE as an optimization which enables packing several signals in only one ciphertext, therefore using a smaller number of ciphertexts or even just one ciphertext, with the corresponding increase in security (higher dimensionalities in the underlying lattices; see Section~\ref{sec:perform}).

For this purpose, we only have to consider one additional polynomial variable that indexes the different messages which are encrypted inside the ciphertext. For example, when dealing with simple images we would use $3$-RLWE (for 3D-images we would resort to $4$-RLWE) in such a way that two polynomial variables would define the content of each image, and the third variable would define the ``index'' of the chosen image (see Figure~\ref{fig:packing}).

It is easy to find further scenarios where this strategy can be applied. For example, when considering the different color layers of the images we can encrypt each layer in a different polynomial variable; hence having a total of $7$ dimensions (R, G, B layers, horizontal and vertical spatial dimensions, the block structure of the images and the number of images). This highlights the versatility of $m$-RLWE.

The difficulty of the implementation in the encrypted domain can vary depending on the operations performed on each block or signal. For example, the computational cost will be smaller or higher depending on whether all the operations are, respectively, the same or different for each block. All the details of the underlying primitives are explained in Section~\ref{sec:scheme}.

\paragraph*{\bf{An example of Block Image Processing}} 

A paradigmatic example of block processing in computer vision can be found in the JPEG compression method, where one step requires to divide the image in blocks of $8 \times 8$ pixels and apply a DCT (Discrete Cosine Transform) to each block.

In~\cite{NTT2015}, the authors describe how to efficiently apply a known linear transform to a signal which has been previously encrypted by a RLWE-based cryptosystem. These techniques can be also applied to this block-wise processing scenario, however the size of needed relinearization matrix would become $2n^2 \lceil \log_t{q} \rceil$ coefficients modulo $q$ (considering a single layer image with $n$ pixels).

Our proposed strategy encodes the block structure with an additional variable. This enables a reduction in the size of the required relinearization matrix, resulting in $128n \lceil \log_t{q} \rceil$ coefficients modulo $q$ for an image with $n$ pixels (we would have to generate the vectors $\bm{a}, \bm{b}$ which are composed of $64\lceil \log_t{q} \rceil$ polynomials with $n$ coefficients).

\subsection{Unattended Encrypted Divisions and Homomorphic Modular Reductions}

A recurrent problem in Secure Signal Processing is the cipher blow-up of the obtained results after several encrypted operations in iterative processes, as a result of the accumulation of the multiplicative factor whenever the encryptions are not ``refreshed'' after each iteration~\cite{TP11}. For mitigating the effect of this overflow we could increase the available space for the encrypted messages (the modulo $t$ in an $m$-RLWE cryptosystem; see Table~\ref{tab:table1}), or consider a homomorphic integer division or quantization after each encrypted iteration (removing the accumulated factor).

In the literature we can find several approaches for computing a secure integer division $\lfloor \frac{a}{b} \rfloor$, but all of them resort to interactive protocols (e.g., \cite{NX10,Veugen14,DNT12}), and they commonly consider that the denominator $b$ in the division is public (\cite{DNT12} keeps it private).

We briefly discuss how to tackle non-interactive encrypted quantizations by resorting to the flexibility of the $m$-RLWE formulation, by including additional (i.e., virtual) polynomial variables. This enables the execution of both real and integer divisions, at the cost of an increase on the cipher expansion.

First, we deal with unattended encrypted integer divisions (always considering that the denominator is public), and then we address how to encode real numbers.

\subsubsection{Integer Divisions}

We can add one variable representing the binary encoding of the different messages (either signal samples or pixels when dealing with images). This implies an increase on the cipher expansion as we encode each value using one polynomial instead of only one coefficient. Thanks to this increase in the cipher expansion (and with the use of binary masks), we enable encrypted integer divisions with a denominator power of $2$.

For performing these integer divisions we can leverage to the tools from \cite{NTT2015}, where we show how to perform shifts and element-wise products between two encrypted messages in an unattended way; the secret key owner only has to generate several relinearization matrices which allow the server to recover the original structure of the ciphertexts after the different operations. Thus, if we work with the binary representation of the different values, we only have to apply a mask which discards the bit(s) with the smallest significance, and afterwards, homomorphically perform the corresponding binary shift.

The efficiency of such scheme is severely limited by the use of a binary decomposition, so we can look for a tradeoff that enhances the performance:
instead of encoding each value using its binary decomposition, we can use a representation in any other base $b>2$. This considerably reduces the cipher expansion while still being able to perform a reduced set of integer divisions by powers of the new base.

Additionally, it is worth noting that the encryption does not hold information about the carries in each position (when they have been previously undergone another homomorphic operations), so the performed divisions could contain errors. To address this, we can adapt the homomorphic threshold function presented in \cite{PTP16} to homomorphically compute the existing carries in each position, therefore correcting the results.

\subsubsection{Working with Real Numbers}

The same additional variable used in the previous paragraphs can be used for a fixed point representation of real numbers. For example, we can use the binary encoding of \cite{DGLLNW15}.
Hence, the polynomial $b_0 + b_1 v + \ldots + b_{N+} v^{N_+} - b_{-1} v^{n_v - 1} - b_{-2} v^{n_v - 2} - \ldots - b_{N_-} v^{n_v - N_-}$ that belongs to the ring $R_2[v] = Z_2[v]/v^{n_v} + 1$ encodes the real number $b_{N_+} \ldots b_1 b_0.b_{-1}b_{-2} \ldots b_{-N_-}$ in base two. After a product of two polynomials encoding two real numbers, if the number of coefficients in the polynomial is big enough for storing the new integer and decimal parts,  we obtain a polynomial that encodes the desired result.

This encoding enables multiplications between real numbers and also real divisions in fixed-point. After an encrypted division between real numbers, we can apply a mask for rounding the corresponding result, hence achieving a better control on the increase of the encrypted values after the homomorphic operations. Analogously, as in the case of integer divisions, we can consider a base $b>2$ for the real fixed point representation.

\subsection{Multi-Scale Approaches}

Both signal processing and computer vision make extensive use of multi-scale representations to work with the content of a signal or image~\cite{Lindeberg94}. In essence, they aim at finding describing structures of the content by means of representing the information as a one-parameter family of smoothed signals which we call the scale-space representation.

Among the most widespread multi-scale approaches, we can highlight pyramids (e.g., Gaussian and Laplacian pyramids) and wavelet transforms (e.g., Gabor and Haar wavelets). In general, both cases require the use of a chain of downsampling and filtering operations. The use of $2$-RLWE to perform wavelet-based operations was introduced in \cite{PTP16}, where we exemplify how to homomorphically perform the denoising of an image in an unattended way. By combining $m$-RLWE-based cryptosystems with the tools introduced in \cite{NTT2015}, which enable the computation of changes on the sampling rate, we can efficiently perform multi-scale processing like wavelet filters and pyramids.

The set of possible applications \cite{MMOP07} enabled by these techniques is really wide and covers some very diverse applications. Among all of them, applications related to medical scenarios are more amenable for the presented solutions, due to their intrinsic privacy constraints. In these scenarios, we can consider several applications dealing with highly sensitive data, like Electrogardiograms - ECG, Electroencephalograms - EEG, Computer Tomography scans, Magnetic Resonance Imaging - MRI, fMRI, among others.

\section{Encrypted Toolset Based on $m$-RLWE}
\label{sec:scheme}
As mentioned above, image processing commonly relies heavily on block-wise processing. This section explains in detail how the block structure of these operations can be incorporated into $m$-RLWE ciphertexts to take advantage of the multivariate structure and the $m$-RLWE formulation. It is worth noting that while we exemplify solutions for image processing scenarios due to their typical block-wise operations, all the results are equally valid and applicable for any scenario dealing with multidimensional signals.

\subsection{Block Processing}
\label{sec:blockprocess}
First, we consider the case where the same processing is applied to each block. The straightforward approach would be to encrypt each block separately and filter each encrypted block independently, effectively considering every block as a different signal. However, we can leverage the $m$-RLWE structure and, instead of encrypting each block separately, we include one additional variable to the encrypted polynomials which assigns one block per coefficient and enables processing different blocks in parallel without separating them (for the case of images that are divided in several blocks, the equivalent would be to use $3$-RLWE for coding the image as a video where each frame is one of the different blocks). That is, incorporating an ``index'' variable to address the block structure, we can work with only one ciphertext for all the blocks or signals.

If we apply under encryption a filter defined in those variables that represent the dimensions of the blocks, we can effectively work with ciphertexts whose underlying lattice dimensionality is much higher than the ciphertexts of the straightfoward approach, so the security is considerably increased. In addition, efficiency is not reduced, as the expansion is not significantly increased, and one encrypted operation is equivalent to processing several blocks in parallel. We address now the case in which each block has to be processed by a different filter.

\subsubsection{New encryption and decryption primitives}
\label{sec:prepostprocess}
When a different filter has to be applied to each block of the multidimensional signal, it is not enough to have one additional variable for coding the pointer to the block structure. This case would be analogous to having a set of independent multidimensional signals, and the corresponding filter has to be applied to each of them. Hence, we need an efficient and secure packing of several independently operable multidimensional signals into only one ciphertext.

To this end, we consider a pre- and post-processing inside the encryption and decryption primitives, respectively, that we explain below, highlighting the differences that have to be accounted for with respect to the univariate primitives of the cryptosystem presented in~\cite{PTP2015}. 

\paragraph{\bf{DFT/IDFT as pre-/post-processing}}
In order to obtain independent blocks, we apply a transform (DFT, Discrete Fourier Transform) along the additional variable defined as the block index. The convolution theorem states that the transform of a cyclic convolution between two signals in the temporal domain is equivalent to the element-wise product of the transforms of the two original signals:

\begin{equation*}
DFT(x[l] \circledast y[l]) = DFT(x[l]) \circ DFT(y[l]) 
\end{equation*}

This means that the operations applied along the variable $l$ will be ``component-wise'' and independent for each coefficient slot.
Hence, we represent the $m$-dimensional signals by means of multivariate polynomials with $m+1$ variables
\begin{equation*}
x(z_1, \ldots, z_m, z) = \sum_{l_1, \ldots, l_m, l}x[l_1, \ldots, l_m, l]z_1^{l_1} \dots z_m^{l_m}z^l,
\end{equation*}
considering $x(\bm{z}, z)$ where $\bm{z} = \left( z_1, \ldots, z_m \right)$ and $\bm{l} = \left( l_1, \ldots, l_m \right)$; $z$ is the variable that indexes the different blocks of $x$, so we compute the DFT with respect to the coefficients (each coefficient represents an $m$-dimensional block) encoded in the variable $z$ (we consider the modular function $1 + z^N$, that is, $N$ blocks). We have the following:
\begin{equation*}
DFT(x[\bm{l}, l]) = \sum_{l = 0}^{N-1} x[\bm{l},l] e^{\frac{-j2 \pi kl}{N}}.
\end{equation*}
If we apply the cyclic convolution (by means of one homomorphic product between ciphertexts) between $\bm{X}[\bm{l}, k]$ and $\bm{H}[\bm{l}, k]$ with respect to the variable $k$, and afterwards the corresponding IDFT with respect to $k$, we are effectively computing the block-wise linear convolution between the blocks that form $x(\bm{z}, z)$ and $h(\bm{z}, z)$ (provided that the results of the linear convolutions do not overflow).

Therefore, if we apply the unidimensional DFT/IDFT across the index variable as pre-/post-processing, we can perform the block-wise linear convolution between all the blocks that form both signals by means of just one homomorphic convolution between $\bm{X}$ and $\bm{H}$.

\paragraph{\bf{Circular Convolution inside the Cryptosystem}}
The correctness of the result of the linear convolution only requires that there be enough coefficients to store it, but the convolution property of the DFT requires a cyclic convolution. It must be noted that the cryptosystem only allows to perform multiplications between polynomials modulo $f(z) = 1 + z^n$ for each variable, so we can only perform nega-cyclic convolutions homomorphically.

Several works (see for example~\cite{Nussbaumer1990}) show how to implement operations modulo $1 + z^n$ by means of cyclic convolutions. Here, we can apply the reverse process (presented in \cite{Murakami00} and generalized in \cite{NTT2015}), for enabling cyclic convolutions using operations between polynomials modulo $f(z) = 1 + z^N$.

First, we have to do a pre-processing before encryption
\begin{align*}
x'[\bm{l}, l] & = x[\bm{l}, l]{(-1)}^{\frac{l}{N}}, \\
h'[\bm{l}, l] & = h[\bm{l}, l]{(-1)}^{\frac{l}{N}},
\end{align*}
for $l = 0, \ldots, N - 1$.

Next, we have to do the post-processing for the resulting $y'(\bm{z}, z) = x'(\bm{z},z)h'(\bm{z},z) \mod 1 + z^N$ after decryption
\begin{equation*}
y[\bm{l}, l] = y'[\bm{l}, l]{(-1)}^{\frac{-l}{N}},
\end{equation*}
for $l = 0, \ldots, N - 1$, and we obtain a homomorphic cyclic convolution.

It is important to note that the presented pre- and post-processing steps require the use of complex numbers to represent the complex roots of $1$ and $-1$. As mentioned in Section~\ref{sec:multi-convolutions}, complex numbers can be accommodated in the used cryptosystem by adding one additional variable with a modular function $f(w) = 1 + w^2$; this effectively doubles the size of the lattices (increasing complexity but also security). 
The main drawback of this solution stems from the need for quantizing the non-integer complex roots represented in fixed-point with sufficient precision; this introduces rounding errors and implies an increase in the needed modulo for representing the signals, therefore increasing also the cipher expansion. In order to remove this constraint and avoid rounding errors, we can replace the DFT by its finite ring counterpart as explained in the next section.

\subsection{Optimizations: Using the NTT to remove rounding errors}

Instead of the complex-valued DFT, we resort to the DFT over finite rings, that is, the NTT (Number Theoretic Transform)~\cite{Nussbaumer1990, NTT2015}. Additionally, we use a finite $N$-th root of $-1$ in $\mathbb{Z}_t$ for the pre- and post-processing of the cyclic convolution. This allows us to avoid both the rounding problems and the need of doubling the size of the used polynomials. This can only be applied for certain values of $t$ and $N$.

The use of the NTT as a method both for efficiently performing encrypted operations and as an encrypted operation inside an RLWE based cryptosystem was introduced by the authors in~\cite{NTT2015}, and exemplified as a pre-/post-processing in~\cite{TPP17} for the univariate case. Hence, here we briefly discuss the particularities of the NTT when applied to the multivariate case, and refer the reader to \cite{NTT2015} for further details.

The existence conditions for an NTT with size $N$ in $\mathbb{Z}_t$ (with $t = \prod_{i=1}^{K} t_i^{m_i}$ where the $t_i$ are different prime numbers) are the following:

\begin{itemize}

\item There exists an $N$-th root of unity $\alpha$ in $\mathbb{Z}_t$ such that $\gcd(\alpha, t) = \gcd(N,t) = 1$.

\item $\gcd((\alpha^i - 1), t) = 1$ for $i = 1, \ldots, N-1$.

\end{itemize}

The expressions for the calculation of the NTT and the INTT are the following:

\begin{equation*} 
\bm{X}[\bm{l}, k] = \sum_{l = 0}^{N-1} x[\bm{l}, l] \alpha^{lk} \mod{p},
\end{equation*}
for $k = 0, 1, \dots, N - 1$ and

\begin{equation*} 
x[\bm{l}, l] = N ^{-1} \sum_{k = 0}^{N-1} \bm{X}[\bm{l}, k] \alpha^{-lk} \mod{p},
\end{equation*}
for $l = 0 , 1, \dots, N - 1$.

It is worth noting that our work focuses on showing the utility of the $m$-RLWE problem for producing efficient and secure solutions that belong to the field of secure signal processing with multidimensional signals. Hence, pre- and post-processing are regarded as a component of the solution, trading-off a slight increase (linear in the size of the input plaintext vectors) in computational cost of the encryption and decryption steps for a global improvement of the security and efficiency of the algorithms. In Section~\ref{sec:perform} we analyze the impact of these pre- and post-processing steps in the computational cost and we show that it is negligible compared with the cost of the (regular) encryption/decryption primitives.

In addition, when the case requires it, it is also possible to offload these pre- and post-processing operations to be performed under encryption (without the intervention of the secret key owner)  by applying the methods proposed in \cite{NTT2015} to the multivariate case, at the cost of an increase in the computational load at the evaluator.

This concludes the basic mechanisms for efficiently operating on $m$-RLWE encryptions. The next section introduces methods to perform on-the-fly changes in the ciphertext structure in an unattended way, which enables homomorphic updates on the available encrypted operations.

\section{Updatable Ciphertext Structure}
\label{sec:switchingkey}
The previous sections show how the possibility of adding some extra structure to the encrypted information together with the use of some pre- and post-processing can enable a unattended encrypted processing in a wide set of practical scenarios. However, once data are encrypted, $m$-RLWE imposes a specific fixed structure optimized for a determined processing, and it is easy to imagine scenarios where the ability to change the underlying ciphertext structure is very convenient (if a chain of processes has to be applied unattendedly).

The straightforward approach would be to send the ciphertext to the secret key owner to decrypt and reencrypt under the new structure. This introduces several problems: a) the user can see some part of the required steps for the execution of the algorithm implemented by the server, and b) this has an increase in the total response time because of the delay caused by the communication between the server and the user. In order to address these two problems, we propose a new mechanism which allows the third party to change the ciphertext structure in an unattended way (without interaction with the secret key owner). To this end, we apply a modification of the \emph{relinearization} procedure~\cite{NTT2015}.

\subsection{Relinearization}

The basic relinearization operation is intended to process encryptions after a homomorphic product. After a product, the encryptions become a function of powers of the secret key $s$. The relinearization builds key homomorphisms that relate $s^2$ to $s$ and is used to produce a 2-component fresh-like encryption from a three-component one. For our purposes, we present a more generic version of the relinearization, which defines key homomorphisms between two keys $s$ and $s'$. Let us consider a ciphertext $(c_0, c_1)$ with decryption circuit $c_0 + c_1s$.
If we apply the relinearization algorithm to $(c_0, c_1)$ to express it as a function of the new key $s'$, we have:

\begin{equation*}
c_0^{relin} = c_0 + \sum_{i = 0}^{\lceil \log_T q \rceil -1} c_{1,i}b_i
\end{equation*}

\begin{equation*}
c_1^{relin} = \sum_{i = 0}^{\lceil \log_T q \rceil -1} c_{1,i}a_i
\end{equation*}

where the set of polynomials $c_{1,i}$ with $i = 0, \ldots, \lceil \log_T q \rceil - 1$ is the base-$T$ decomposition of $c_1$ for a given $0<T<q$.\footnote{We assume that $T=t$ unless otherwise stated.} The different $b_i$ and $a_i$ come from the key homomorphism $h_i = (a_i, b_i = -(s'a_i + Te_i) + T^is)$ with $i = 0, \ldots, \lceil \log_T q \rceil - 1$; these homomorphisms can be seen as ``pseudoencryptions'' of the key $s$ under $s'$. For the sake of exposition, the decryption circuit of $(c_0, c_1)$ can be represented in matrix notation as $\boldsymbol{c}_0 + \boldsymbol{C}_1\boldsymbol{s}$, where $\boldsymbol{C}_1$ is a block skew circulant matrix of the polynomial $c_1$ \cite{NTT2015}. The matrix notation allows to see the decryption equation as a sum of external products of restructured versions of the polynomial $c_1$ times each of the coefficients of the key: $c_0 + \sum_{j = 0}^{n-1} c^{(j)}_1s_j$ where the different $c^{(j)}_1$ are polynomials whose coefficients are the elements of the $j$-th column of the skew circulant matrix $\boldsymbol{C}_1$. 
In general, if we consider the concatenation of $n$ key homomorphisms $h_i^{(j)}$ with $i = 0, \ldots, \lceil \log_T q \rceil - 1$ and $j = 0, \ldots, n - 1$, where $h_i^{(j)}$ has the coefficient $s_j$ ``pseudo-encrypted'' with the secret key $s'$, we can obtain a new ciphertext $(c_0^{relin}, c_1^{relin})$ without changing its content.

\subsection{Changing the polynomial structure by resorting to the relinearization process}

The introduced representation of the decryption circuit ($c_0 + \sum_{j = 0}^{n-1} c^{(j)}_1s_j$) already sheds some light about the approach we follow to change the polynomial structure through a relinearization operation: we simply encode the different polynomials that form the $h_i^{(j)}$ along with $c_0$ and $c^{(j)}_1$ under the desired polynomial structure.

In order to incorporate this new structure, we first define a family of $n!$ different reversible polynomial ring mappings $f_{\bm{n},\bm{m}}^{(w)}:  R_q[z_1, \ldots, z_l] \rightarrow R_q[x_1, \ldots, x_k]$ with $w$ belonging to the set $\{1, \ldots, n!\}$ where $\bm{n} = (n_1, \ldots, n_l)$, $\bm{m} = (m_1, \ldots, m_k)$ and $n = \prod_{i = 1}^{l} n_i = \prod_{i = 1}^{k} m_i$ (the modular functions of the polynomial rings are $f_i(z_i) = z_i^{n_i} + 1$ with $i = 1, \ldots, l$, and $f_j(x_j) = x_j^{m_j} + 1$ with $j = 1, \ldots, k$).

This mapping takes as input a polynomial element that belongs to the ring $R_q[z_1, \ldots, z_l]$ and produces as output a polynomial element that belongs to the ring $R_q[x_1, \ldots, x_k]$ and whose coefficients are the same as the coefficients of the polynomial input but rearranged in one of the $n!$ different ways ($w$ indicates the specific reordering used).

Now, we need a set of key homomorphisms $h_i^{(j)}$ with $j = 0, \ldots, n-1$ where all the used polynomials belong to the output polynomial ring, that is $a_i,e_i \leftarrow R_q[x_1, \ldots, x_k]$, and where instead of using $s \in R_q[z_1, \ldots, z_l]$ we are ``pseudo-encrypting'' the coefficients $s_j$ with the secret key $f_{\bm{n},\bm{m}}^{(w)}(s) \in R_q[x_1, \ldots, x_k]$.

Equipped with these tools, we perform a relinearization in which we consider the use of $f_{\bm{n},\bm{m}}^{(w)}(c_0)$, $f_{\bm{n},\bm{m}}^{(w)}(c_1^{(j)})$ for $j = 0, \ldots, n-1$ instead of $c_0$ and $c_1^{(j)}$. By doing this, we obtain a new ciphertext $(c_0^{relin}, c_1^{relin})$ that is the encryption of $f_{\bm{n},\bm{m}}^{(w)}(m) \in R_t[x_1, \ldots, x_k]$ (the corresponding reordering of the original message $m \in R_t[z_1, \ldots, z_l]$) with the secret key $f_{\bm{n},\bm{m}}^{(w)}(s) \in R_q[x_1, \ldots, x_k]$ and where $c_0^{relin}, c_1^{relin} \in R_q[x_1, \ldots, x_k]$. 

For example, if both $c_0$ and $c_1$ are polynomials that belong to $\mathbb{Z}_q[z]/(1 + z^n)$ and we want to divide the encrypted signal in blocks of length $n_x$ (e.g., to obtain an image whose rows are the different blocks), we consider the ring $( \mathbb{Z}_q[x, y]/(1 + x^{n_x}) ) /(1 + y^{n_y})$ with $n_xn_y = n$; being $n_x$ and $n_y$ powers of $2$. As we know which is the new position of each coefficient of the encrypted message in the new multivariate structure, we apply the explained method considering that the polynomials belong to the bivariate ring $(\mathbb{Z}_q[x, y]/(1 + x^{n_x}))/(1 + y^{n_y})$.

The presented strategy can be applied to change the structure of the encrypted messages to all types of multivariate polynomials depending on what we need.

\paragraph*{\bf{Security considerations}}
The security of this process is guaranteed by the underlying $m$-RLWE problems (see Section~\ref{sec:mrlwe}) involved in the execution of the algorithm. Consider that we have a chain of structure changes defined by a composition of $L$ mappings $f^{(w_1)}_{\bm{n}^{(1)},\bm{n}^{(2)}} \circ f^{(w_2)}_{\bm{n}^{(2)},\bm{n}^{(3)}} \circ \ldots \circ f^{(w_{L})}_{\bm{n}^{(L)},\bm{n}^{(L + 1)}}$, where each $w_i$ belongs to the set $\{1, \ldots, n!\}$ with $i = 1, \ldots, L$ and each $\bm{n}^{(j)} = \left( n^{(j)}_1, \ldots, n^{(j)}_{k_j} \right)$ with $j = 1, \ldots, L + 1$ is a vector composed of $k_j$ natural numbers satisfying $n = \prod_{i = 1}^{k_1} n^{(1)}_i = \prod_{i = 1}^{k_2} n^{(2)}_i = \ldots = \prod_{i = 1}^{k_{L + 1}} n^{(L + 1)}_i$. Then, the security of the proposed algorithm is based on the hardness of the underlying multivariate RLWE problems defined over the $L + 1$ rings $R_q[z^{(j)}_1, \ldots, z^{(j)}_{k_j}]$, where the different modular functions are defined as in the previous section, that is, $f_{k_j}(z^{(j)}_{k_j}) = {(z^{(j)}_{k_j})}^{n^{(j)}_{k_j}} + 1$. Additionally, the security is also based on the circular security of the different involved multivariate RLWE based cryptosystems (see Section~\ref{sec:applic}), hence guaranteeing that releasing encryptions of the secret key is secure.

\section{Performance Evaluation}
\label{sec:perform}
In this section, we compare our implementation of the proposed primitives and tools with Paillier and RLWE-based approaches in terms of efficiency (computational cost and runtime) and security.
For the latter, we consider distinguishing attacks, that are the best known attacks against lattice-based cryptosystems (i.e., attacks whose objective is to break the indistinguishability assumption by means of basis reduction algorithms). The considered security parameter is the root Hermite factor $\delta$ (the runtime of the attack is approximately propotional to $e^{\frac{K}{\log_2 \delta}}$), which allows to estimate the bit security as~\cite{LNV11, NTT2015, LP11}:
\begin{align}
\label{eq:delta}
\log_{2}(\delta) = {(\log_{2}(c \cdot q/s))}^{2}/(4 n \log_{2}(q)),\; c \approx \sqrt {\ln (\frac{1}{\epsilon})/\pi} \\\label{eq:bit-sec-rlwe}
t_{BKZ}(\delta) = \log_2{\left( T_{BKZ}(\delta) \right)} = \frac{1.8}{\log_2{\delta}} - 110,
\end{align}
where $\epsilon$ is the attacker's advantage.

Due to space constraints, we do not explicitly tackle decoding attacks (whose objective is to obtain the secret key of the cryptosystem), but we use values for $n$ that achieve appropriate protection against the decoding attacks described in \cite{PTP2015}, \cite{LNV11} and \cite{LP11}.

For the comparison, we have chosen an image processing scenario, and more specifically, parallel processing of several images and block image processing, described in the next subsection. The rest of the section analyzes the impact of our proposed schemes on encryption and decryption, and compares the complexity, runtimes and achieved security with respect to previous approaches.

\subsection{Evaluation for Encrypted Image Processing}
We consider two privacy-aware ubiquitous scenarios of outsourced image processing: encrypted correlation between $I$ pairs of square images of size $N \times N$, and encrypted filtering among $I$ pairs of square $N \times N$ images and filters of size $F \times F$ (where $F < N$); the latter is equivalent to block image processing with $I$ blocks of size $N \times N$.

In all cases, we compare the security and efficiency of our proposed multivariate approaches against the univariate and bivariate cryptosystems from \cite{LNV11} and \cite{PTP2015}. The parameters for our cryptosystems are the following: $s = 2\sqrt{n}\Rightarrow\sigma=s/\sqrt{2\pi}$, $t = 257$ (8-bit images), $A = 1$ and $D = 1$ (see Section~\ref{sec:applic}); for a fair comparison, we use a slack variable $h$ for tuning the value of $\delta$ \cite{PTP2015} and comparing the cryptosystems in terms of equal efficiency or equal security level. This slack variable represents the ratio between the total degree $n = \prod n_i$ needed for achieving a certain level of security, and the minimum length needed for storing the result in the different ciphered dimensions.

\subsubsection{Encrypted correlation of a set of images}
In order to fit the result when correlating images of size $N \times N$, the minimum degree of the used polynomials must be: $n \geq 2N - 1$ for Lauter~\cite{LNV11} (RLWE); $n_i \geq 2N - 1$ with $i = 1, 2$ for~\cite{PTP2015} ($2$-RLWE), and $n_i \geq 2N - 1$, with $i = 1, 2$ and $n_3 = I$ for the proposed scheme (in this case, $3$-RLWE). This is the minimum value for the degree, but in order to provide a fair comparison, we account for the aforementioned slack variables $h$ for fixing the same $\delta$ across all the three cases. The approximate relations for the $\delta$ and the computational cost (in terms of number of polynomial products) in the $3$ cryptosytems are the following:
\begin{equation*}
\begin{split}
\mbox{Cost}_{3-RLWE} \approx & I \frac{h_{3-RLWE}^2}{h_{2-RLWE}^2} \mbox{Cost}_{2-RLWE} \\
                     \approx & 4 \left( I \frac{h_{3-RLWE}^2}{h_{2-RLWE}^2} \right) \left( \frac{h_{2-RLWE}^2}{h_{RLWE}^2} \right)\mbox{Cost}_{RLWE} \\
\log_2{\delta_{3-RLWE}} \approx & \frac{h_{2-RLWE}}{h_{3-RLWE}I}\log_2{\delta_{2-RLWE}} \\
                                                 \approx & \left( \frac{h_{RLWE}}{h_{3-RLWE}} \right)\frac{\log_2{\delta_{RLWE}}}{(2N-1)I}.
\end{split}                                                 
\end{equation*}

\begin{figure}
\centering
\subfigure[$I = 8$]{
  \centering
  \includegraphics[width=3.44in]{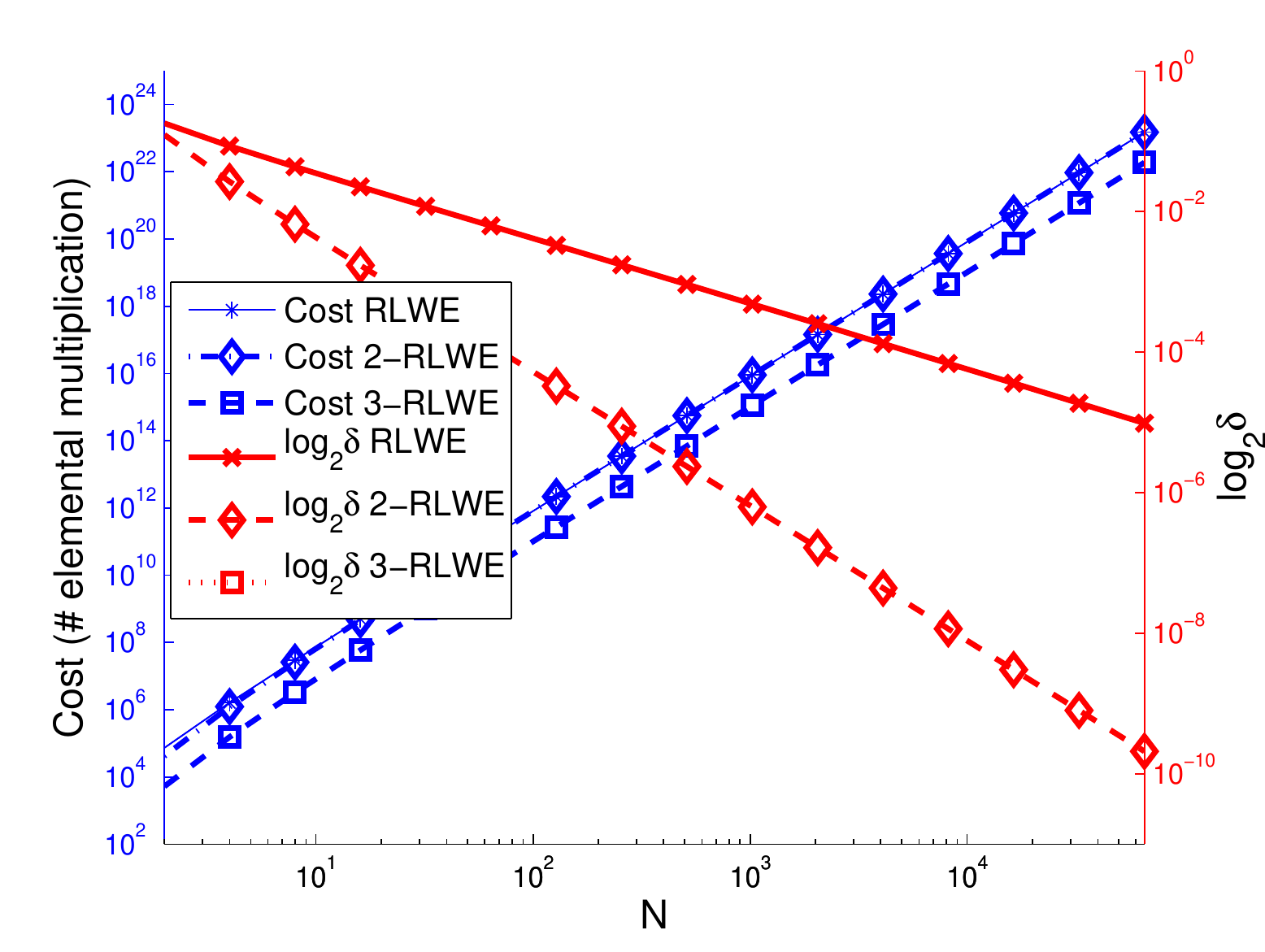}
}
\subfigure[$I = 16$]{
  \centering
  \includegraphics[width=3.44in]{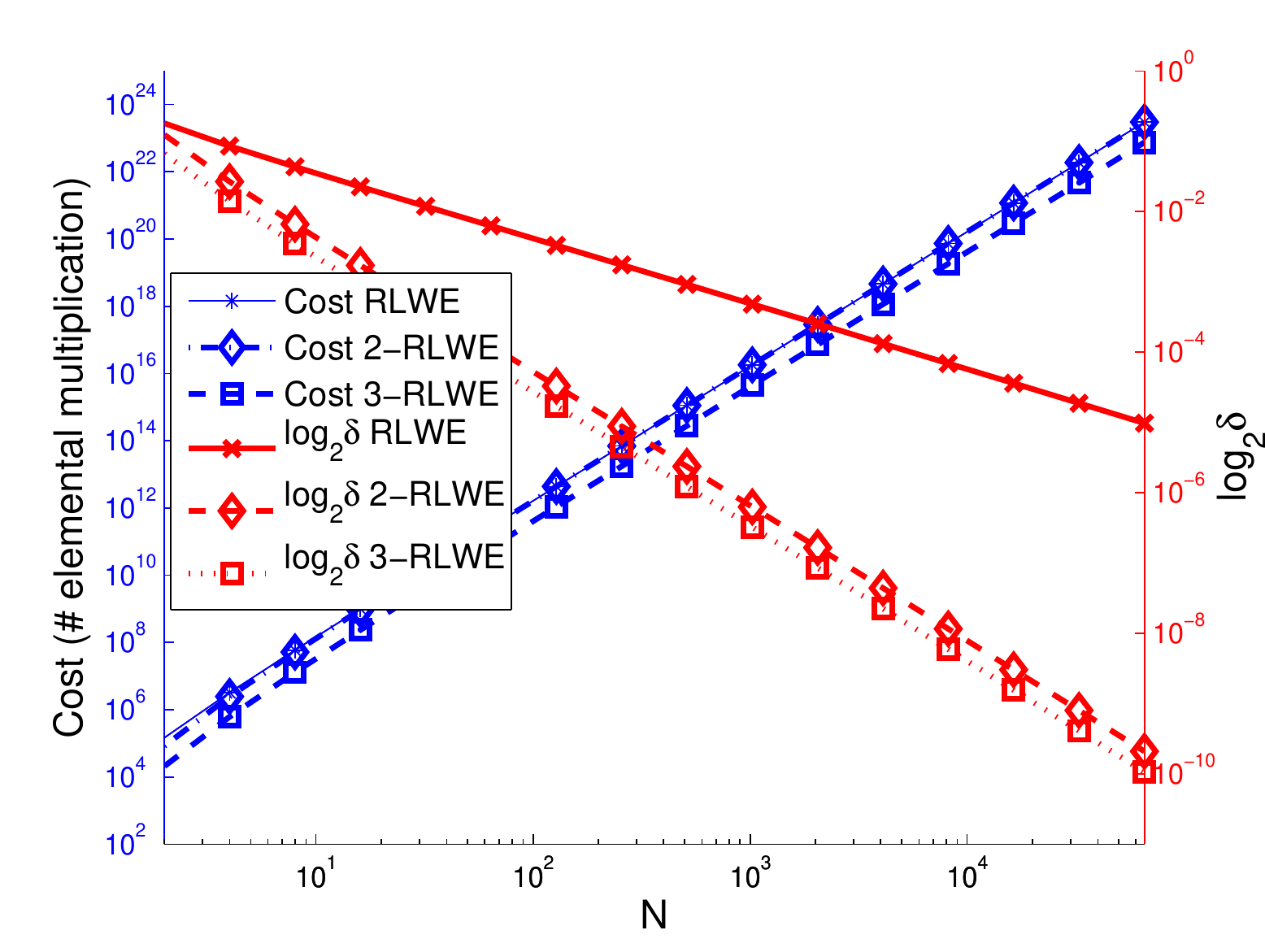}
}\\

\subfigure[$I = 32$]{
  \centering
  \includegraphics[width=3.44in]{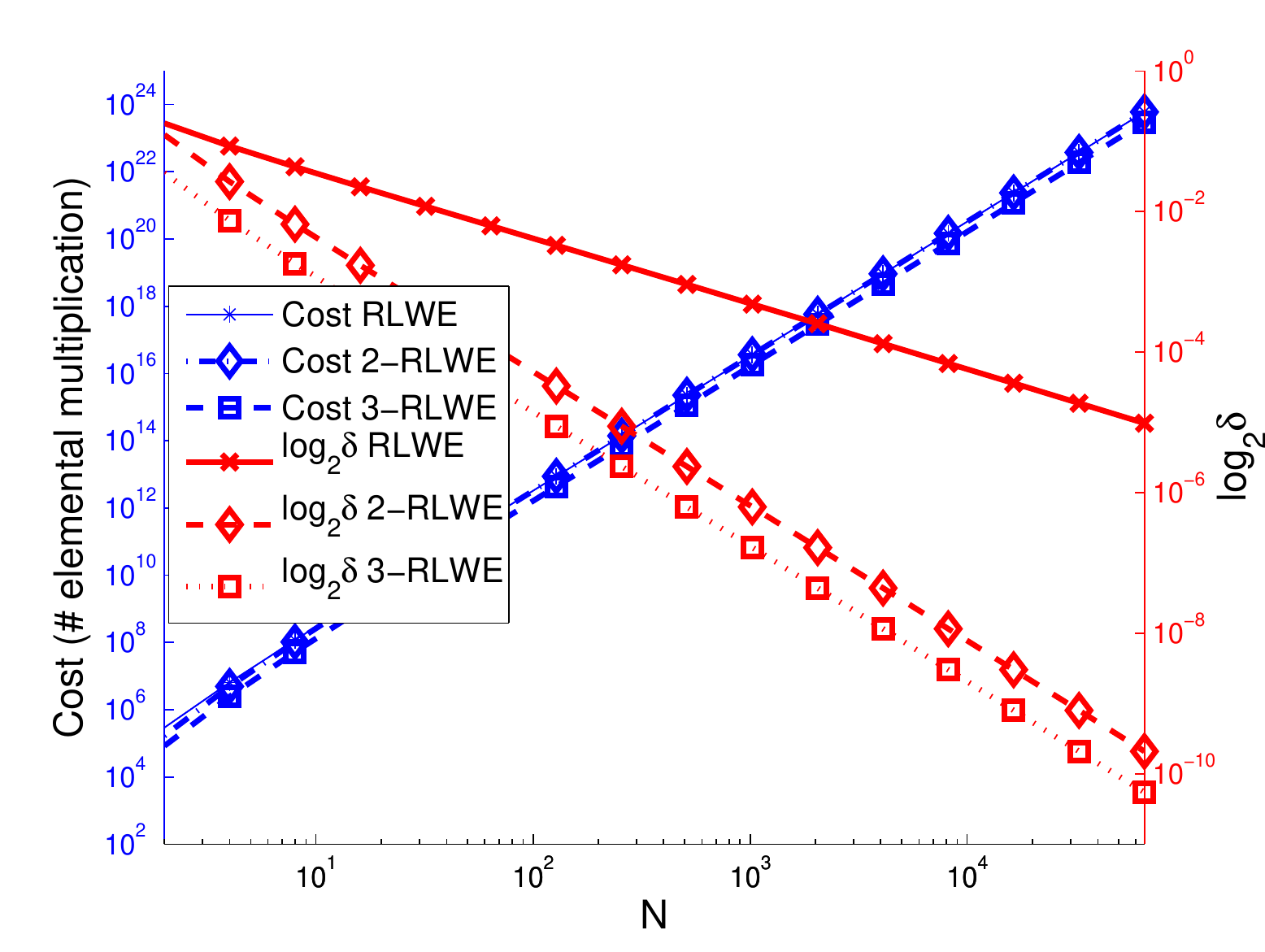}
}
\subfigure[$I = 64$]{
  \centering
  \includegraphics[width=3.44in]{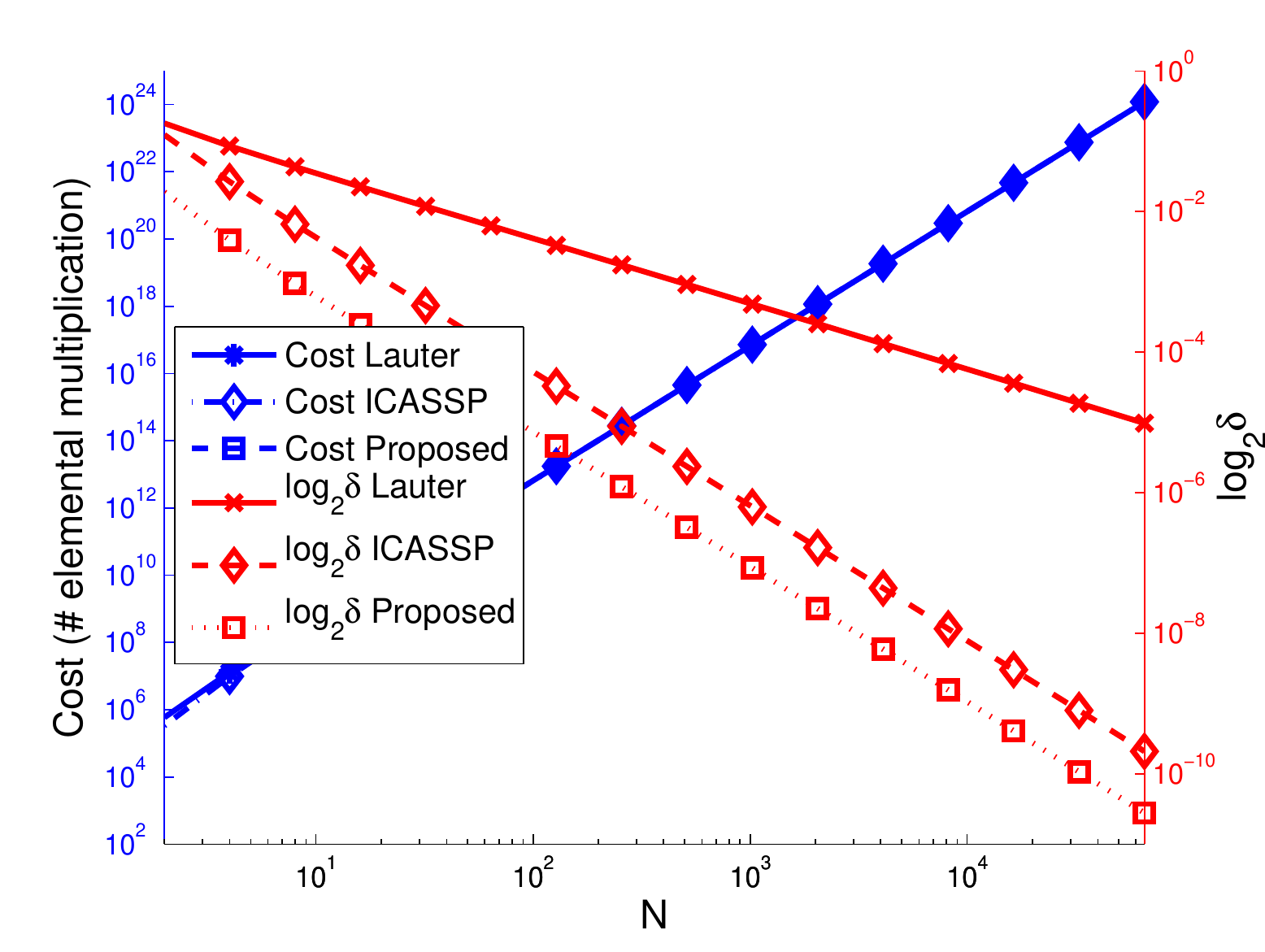}
}\\

\caption{Encrypted image correlation: Performance and security comparison}
\label{fig:cor}
\end{figure}

Table~\ref{tab:seccompcost} reports the comparison in terms of encrypted image size, used polynomial degree and number of ciphertext products for each cryptosystem as a function of the slack variables, while Figure~\ref{fig:cor} shows the computational cost and achieved security with $h_{RLWE} = 16$, $h_{2-RLWE} = 8$ and $I = 8, 16, 32, 64$ varying $N$ (size of the correlated images).

\begin{table}[!t]
\renewcommand{\arraystretch}{1.3}
\caption{Size and performance comparison for encrypted image correlation}
\label{tab:seccompcost}
\centering \tiny
\begin{tabular}{|c|c|c|}
\hline
 Lauter~\cite{LNV11} (RLWE) & $2$-RLWE~\cite{PTP2015} & $3$-RLWE \\
\hline
\hline 
\multicolumn{3}{|c|}{$n$}\\
\hline
 $(2N - 1)h_{RLWE}$ & ${(2N - 1)}^2h_{2-RLWE}$ & ${(2N - 1)}^2h_{3-RLWE}I$ \\
\hline
\hline 
\multicolumn{3}{|c|}{number of ciphertexts}\\
\hline 
 $2NI$ & $2I$ & $2$ \\
\hline
\hline 
\multicolumn{3}{|c|}{ciphertexts products}\\
\hline
 $N^2I$ & $I$ & $1$ \\
\hline
\end{tabular}
\end{table}

It can be seen that for the same level of security ($I = 8$), our approach shows better performance (lower cost), while for the same computational cost ($I = 64$) it is more secure; finally, our approach is both more secure and efficient for intermediate values of $I$ ($I = 16, 32$).

\subsubsection{Encrypted filtering of a set of images}
Our second use case deals with encrypted filtering of several $N\times N$ images with filters of size $F\times F$. In this case, the minimum degree of the used polynomials is: $n \geq N + F - 1$ for Lauter~\cite{LNV11} (RLWE); $n_i \geq N + F - 1$, with $i = 1, 2$ for~\cite{PTP2015} ($2$-RLWE), and $n_i \geq N + F - 1$, with $i = 1, 2$ and $n_3 = I$ for the proposed scheme (in this case, $3$-RLWE). As for the correlation scenario, we have to use slack variables in the Lauter and \cite{PTP2015} cryptosystems in order to guarantee a certain value of $\delta$.
Assuming $F \ll N$, we can approximate the relations between security and computational cost for the three cryptosystems as
\begin{equation*}
\begin{split}
\mbox{Cost}_{3-RLWE} \approx & I \frac{h_{3-RLWE}^2}{h_{2-RLWE}^2} \mbox{Cost}_{2-RLWE} \\
                     \approx & \left( I \frac{h_{3-RLWE}^2}{h_{2-RLWE}^2} \right) \left( \frac{h_{2-RLWE}^2}{h_{RLWE}^2} \right) \frac{N}{F} \mbox{Cost}_{RLWE} \\
\log_2{\delta_{3-RLWE}} \approx & \frac{h_{2-RLWE}}{h_{3-RLWE}I}\log_2{\delta_{2-RLWE}} \\
                                                 \approx & \left( \frac{h_{RLWE}}{h_{3-RLWE}} \right)\frac{\log_2{\delta_{RLWE}}}{(N + F -1)I}
\end{split}                                                 
\end{equation*}

\begin{figure} [ht!]
\centering
\subfigure[$I = 8$]{
  \centering
  \includegraphics[width=3.44in]{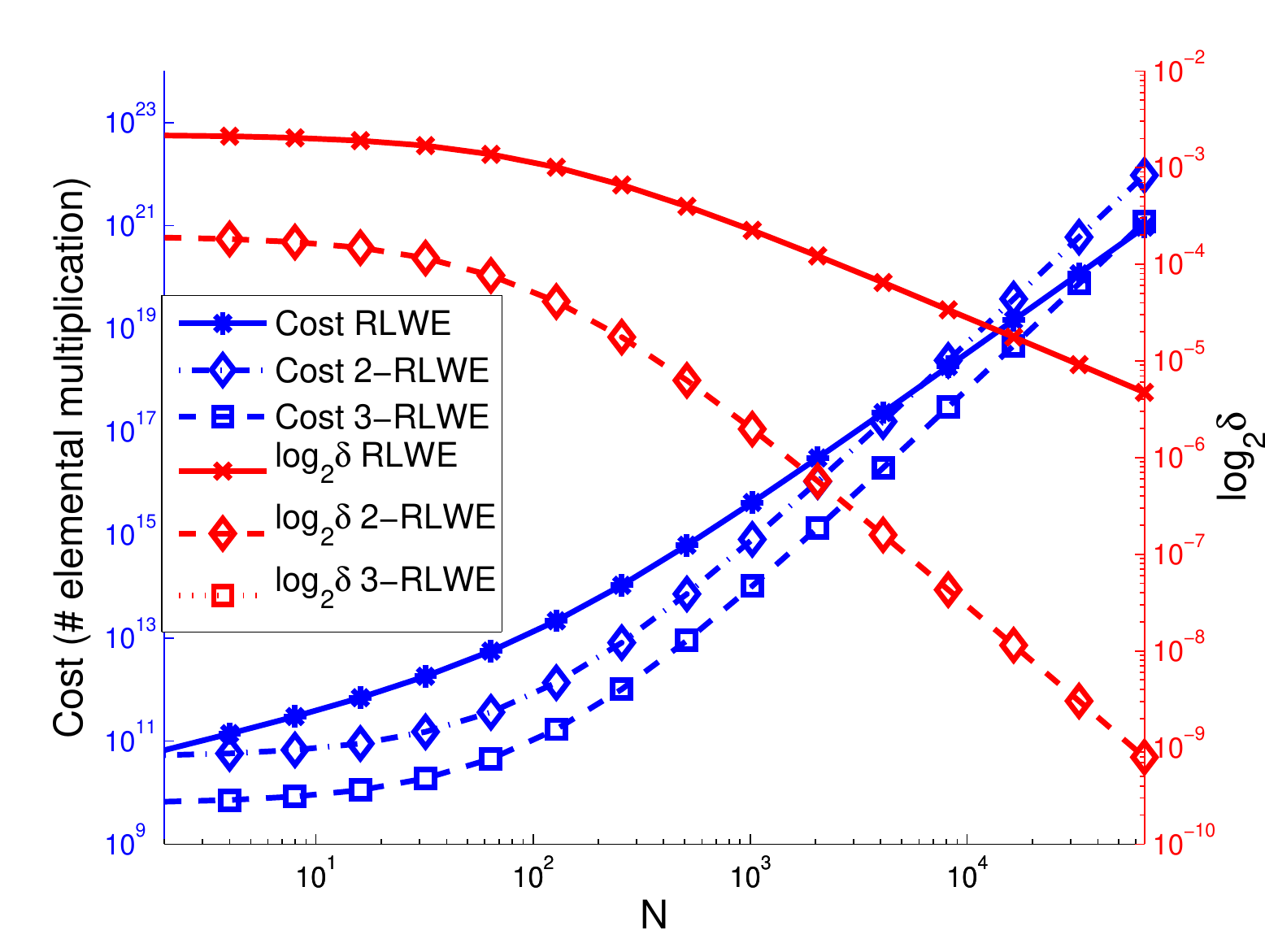}
  \label{fig:sfig1}
}
\subfigure[$I = 16$]{
  \centering
  \includegraphics[width=3.44in]{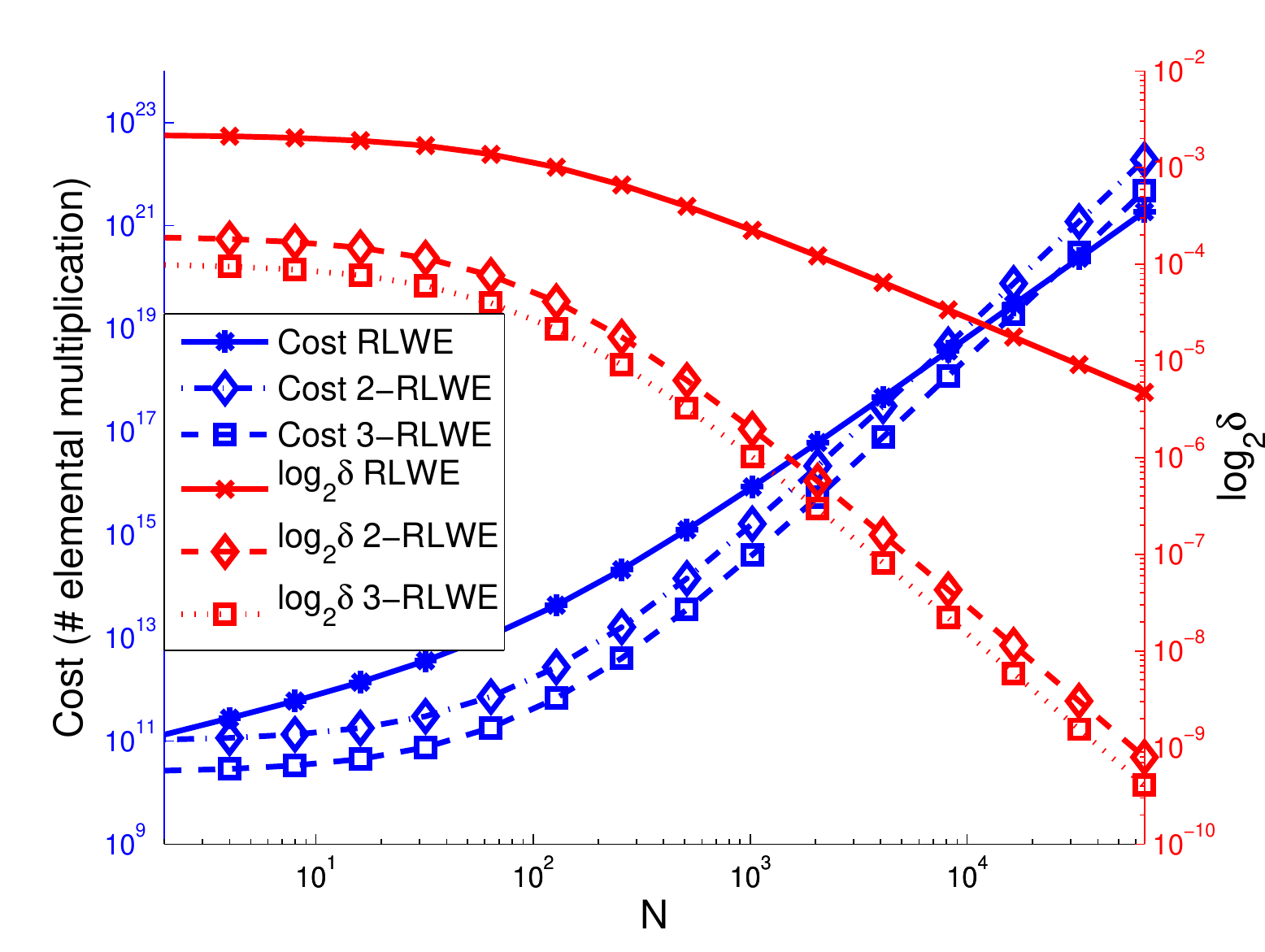}
}\\

\subfigure[$I = 32$]{
  \centering
  \includegraphics[width=3.44in]{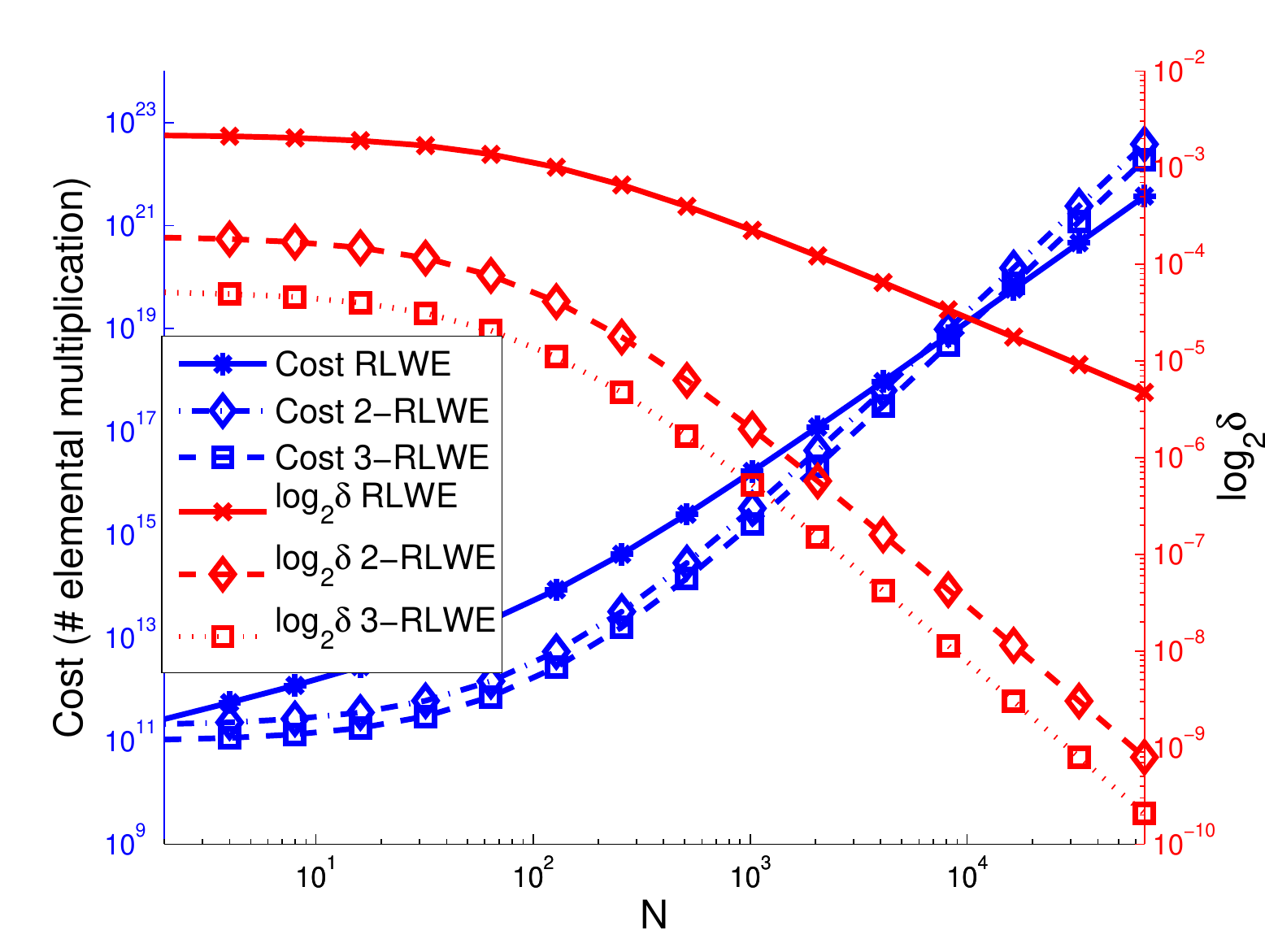}
}
\subfigure[$I = 64$]{
  \centering
  \includegraphics[width=3.44in]{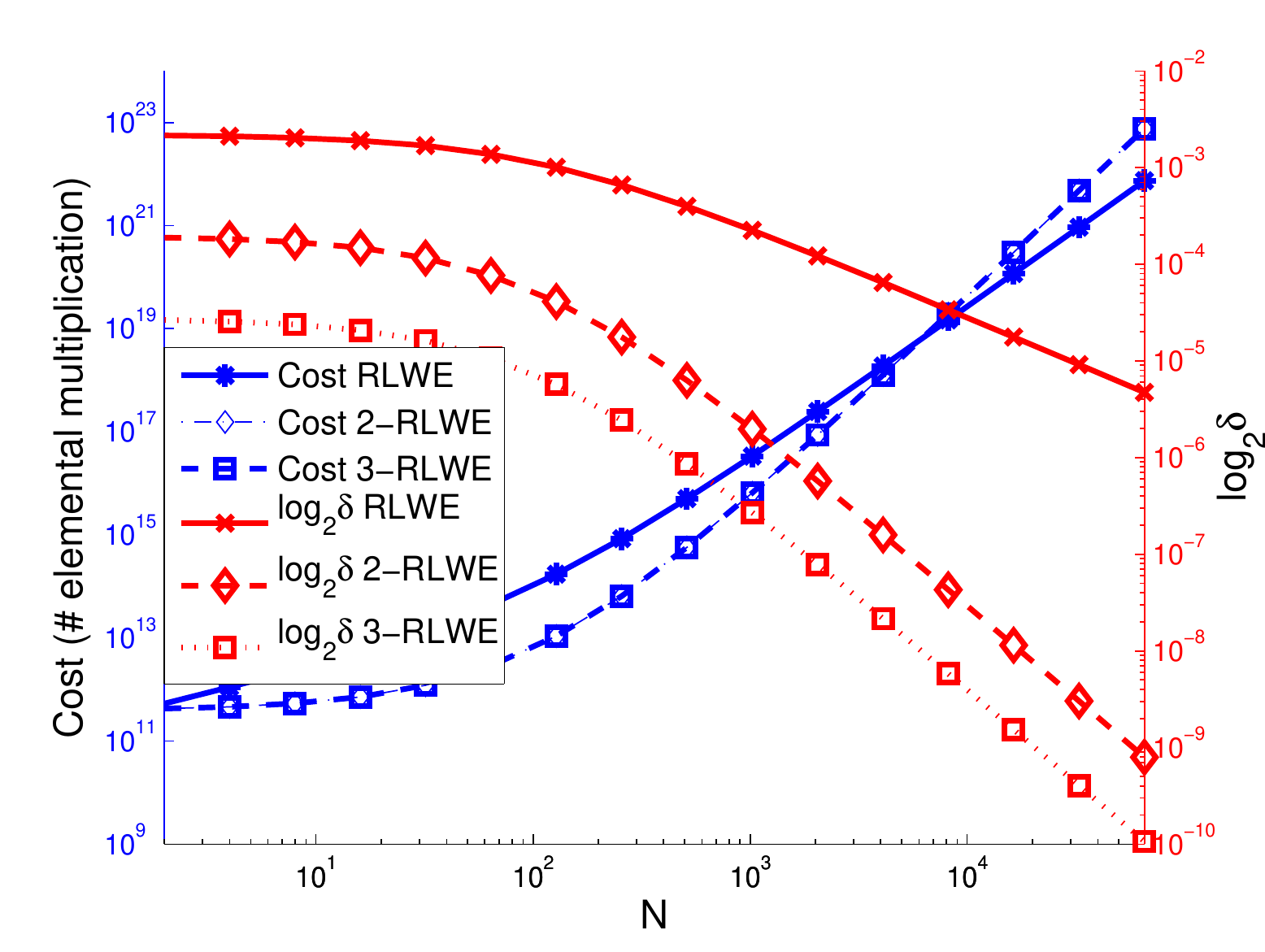}
}\\

\caption{Encrypted image filtering: Performance/security as a function of $N$ (image size)}
\label{fig:fil}
\end{figure}

\begin{figure} [ht!]
\centering
\subfigure[$h_{RLWE} = 32$]{
  \centering
  \includegraphics[width=3.44in]{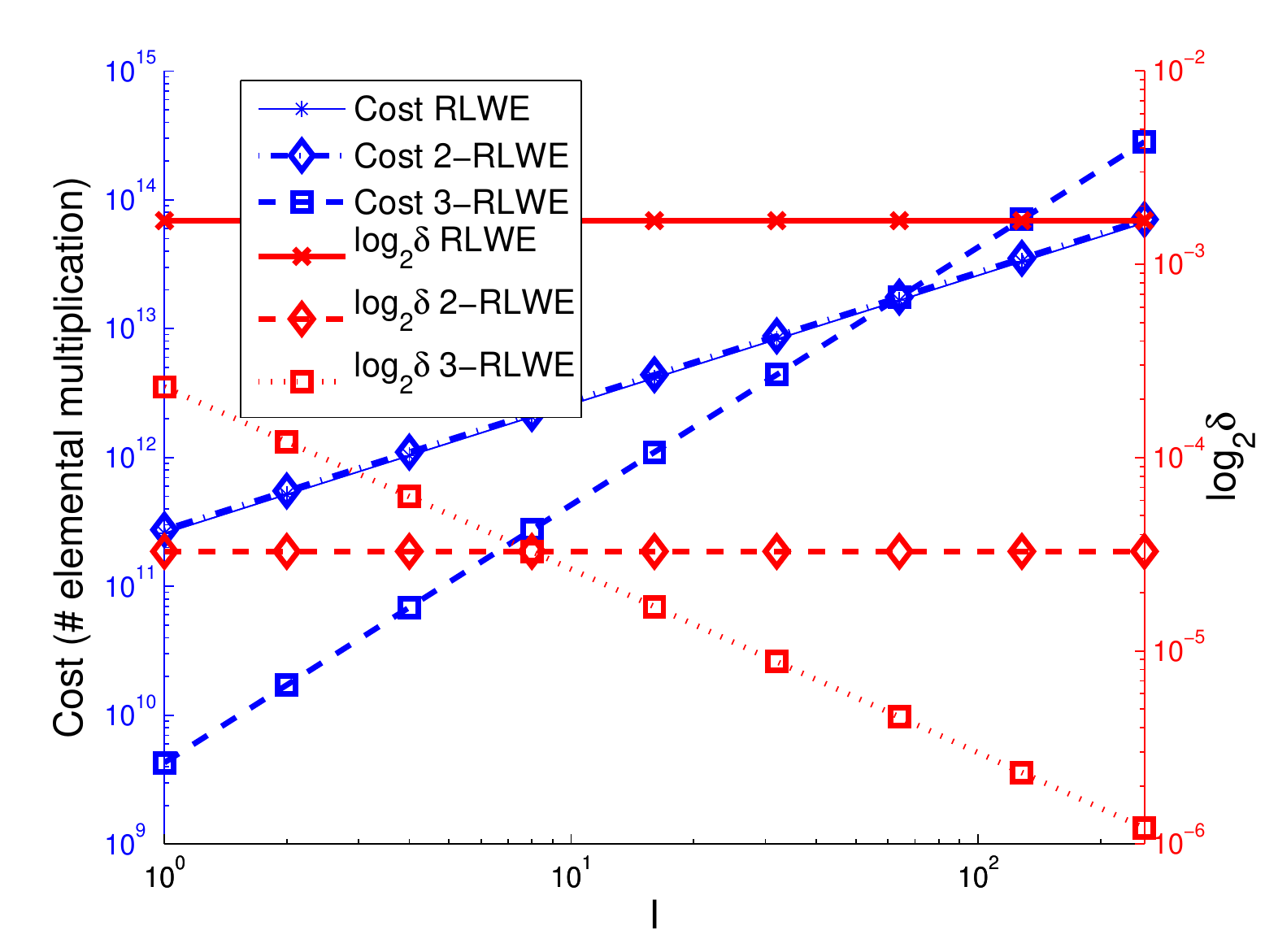}
  \label{fig:sfig2}
}
\subfigure[$h_{RLWE} = 64$]{
  \centering
  \includegraphics[width=3.44in]{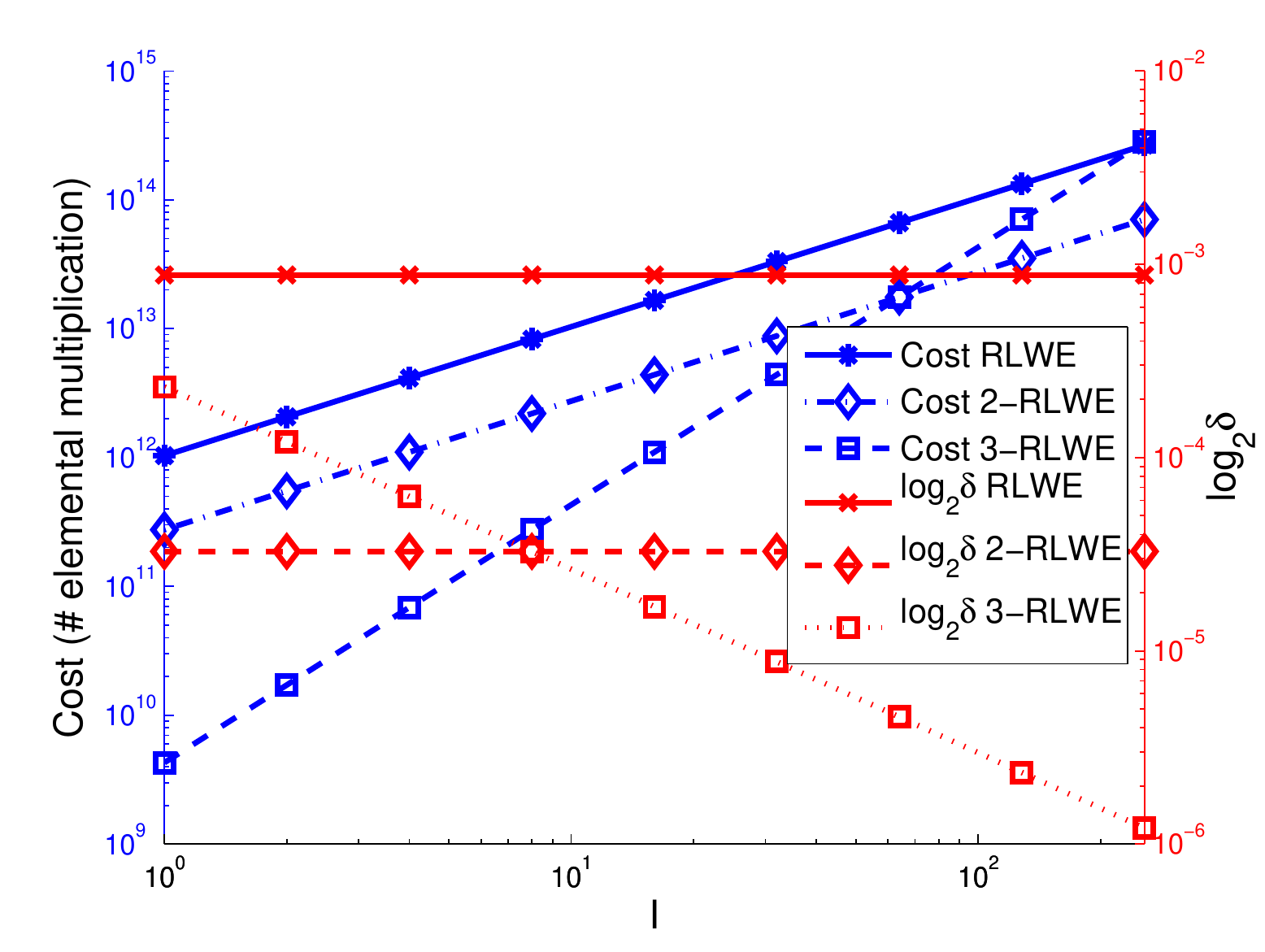}
}\\

\caption{Encrypted image filtering: Performance/security as a function of $I$ (\# of blocks)}
\label{fig:varI}
\end{figure}

\begin{figure} [ht!]
\centering
\subfigure[$I = 8$]{
  \centering
  \includegraphics[width=3.44in]{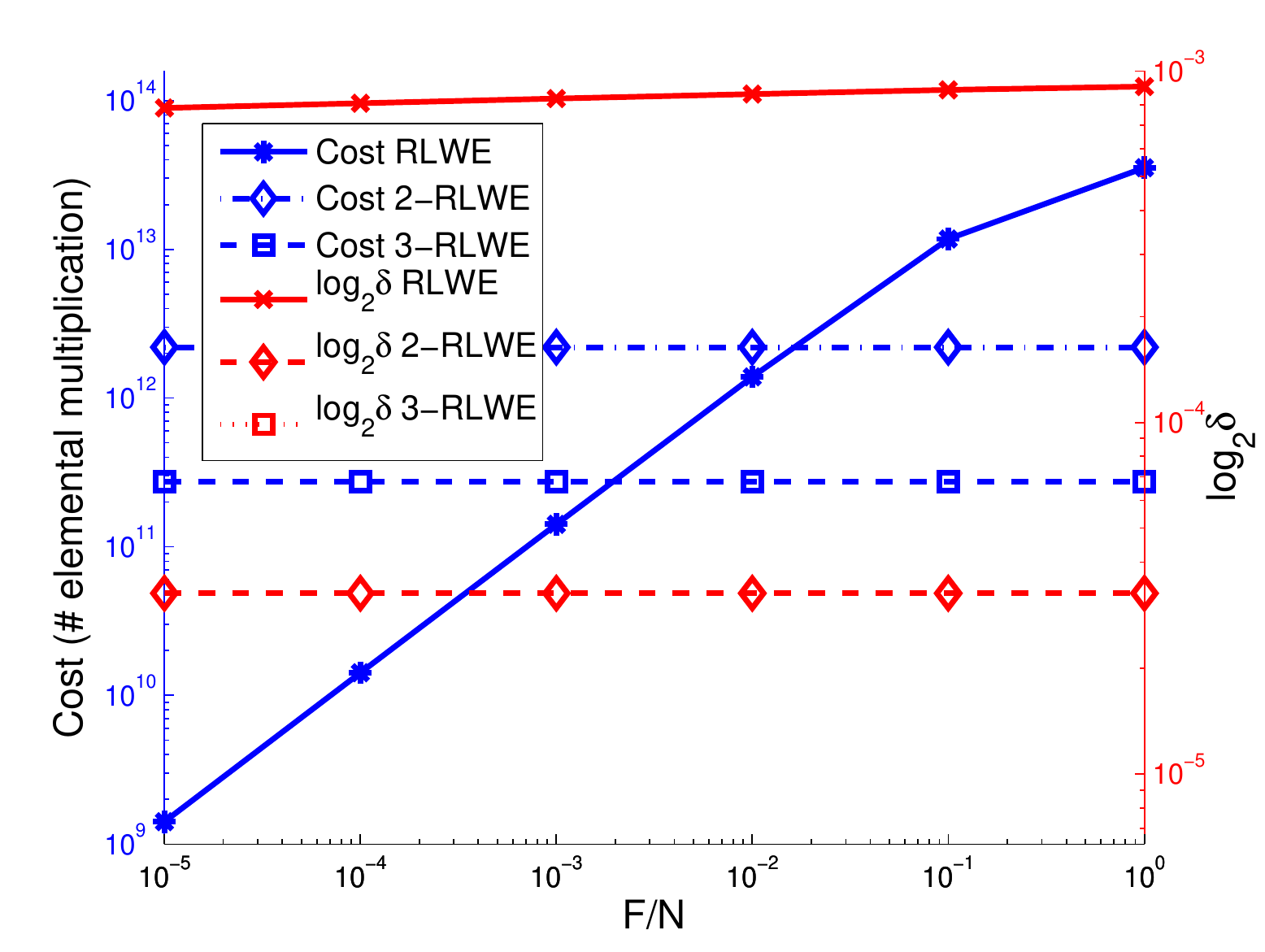}
  \label{fig:sfig3}
}
\subfigure[$I = 32$]{
  \centering
  \includegraphics[width=3.44in]{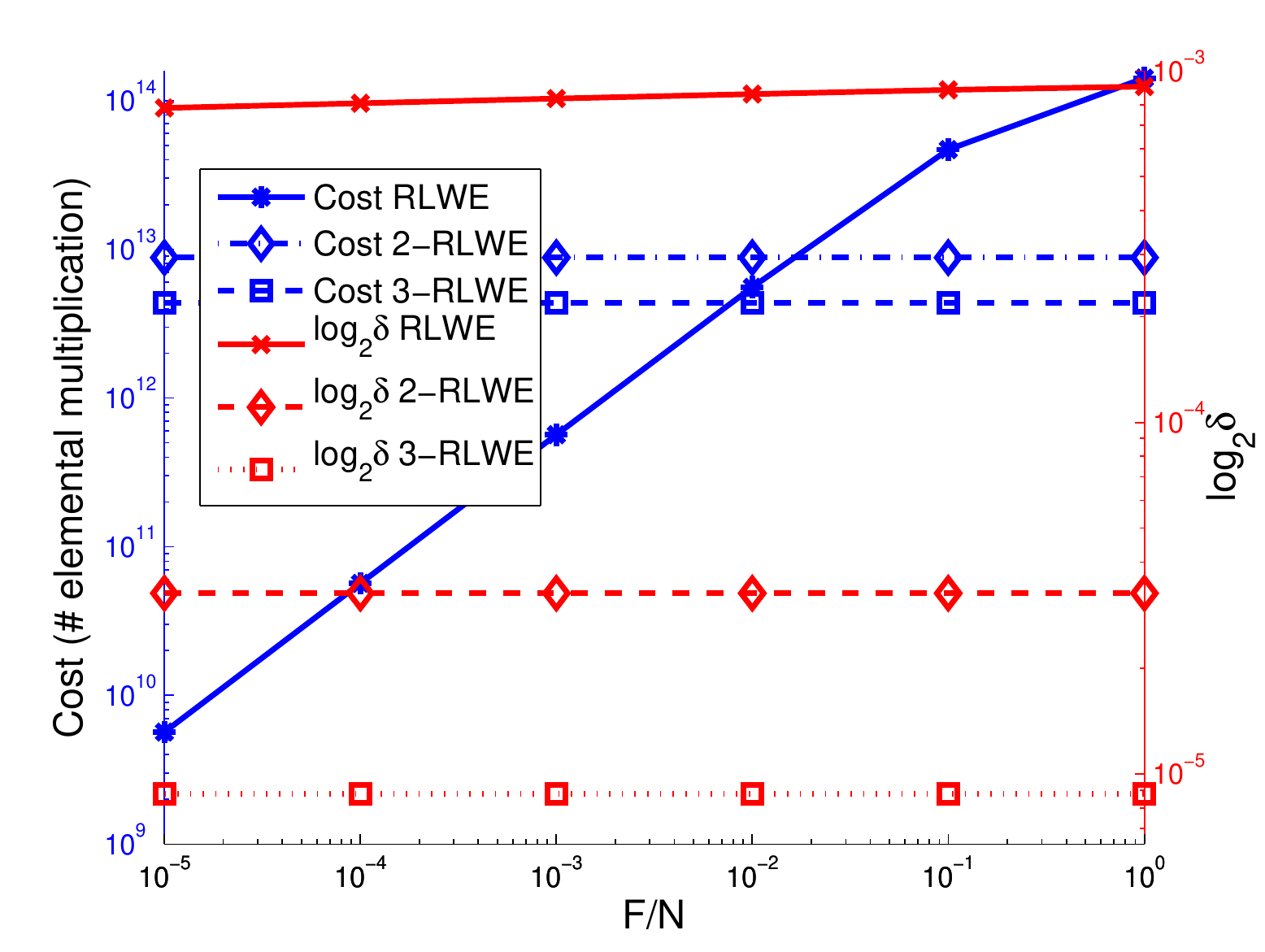}
}\\

\caption{Encrypted image filtering varying the ratio $F/N$}
\label{fig:ratiofN}
\end{figure}

Table~\ref{tab:table2} reports the comparison in terms of encrypted image and filter size, polynomial degree and number of ciphertext products for each cryptosystem as a function of the slack variables, while Figures~\ref{fig:fil}-\ref{fig:ratiofN} compare performance and security for several parameters. Specifically, Figure~\ref{fig:fil} shows the effect of varying $N$ and fixed $F = 100$, $h_{RLWE} = 64$, $h_{2-RLWE} = 8$, for a different number of block operations $I = 8, 16, 32, 64$. The performance and security of our method improves with respect to the univariate and bivariate cases for a wide range of $N$. Figure \ref{fig:varI} varies only the number of images $I$, while fixing the rest of parameters to $N = 241$, $F = 16$, $h_{RLWE} = 32, 64$, and $h_{2-RLWE} = 8$. This figure shows that our approach is more sensitive to increases in the number of packed images or blocks $I$ in terms of cost, but the counterpart is that the underlying security is also considerably increased instead of constant. Finally, Figure \ref{fig:ratiofN} varies the ratio of filter size vs image size $F/N$ while fixing the size of lattices as $N + F - 1 = 256$, with $h_{RLWE} = 64$, $h_{2-RLWE} = 8$ and $I = 8, 32$. We can clearly see that the unidimensional cryptosystem can become more efficient for very small filters, but at the expense of a much smaller security than with $2$-RLWE and $3$-RLWE. For medium-sized filters, our approach becomes again the most efficient and secure.

\begin{table}[!th]
\renewcommand{\arraystretch}{1.3}
\caption{Size and performance comparison for encrypted image filtering}
\label{tab:table2}
\centering \tiny
\begin{tabular}{|c|c|c|}
\hline
 Lauter~\cite{LNV11} (RLWE) & $2$-RLWE~\cite{PTP2015} & $3$-RLWE \\
\hline
\hline 
\multicolumn{3}{|c|}{$n$}\\
\hline
 $(N + F - 1)h_{RLWE}$ & ${(N + F - 1)}^2h_{2-RLWE}$ & ${(N + F - 1)}^2h_{3-RLWE}I$ \\
\hline
\hline 
\multicolumn{3}{|c|}{number of ciphertexts}\\
\hline
 $(N + F)I$ & $2I$ & $2$ \\
\hline 
\hline
\multicolumn{3}{|c|}{ciphertexts products}\\
\hline
 $NFI$ & $I$ & $1$ \\
\hline
\end{tabular}
\end{table}

In summary, by packing several images into one ciphertext, we can achieve both a higher security and a higher eficiency for a wide set of parameter ranges. Depending on the actual application, we can always get the best efficiency by relying on a combination of a $3$-RLWE and a $2$-RLWE cryptosystem, that is, by searching for the optimum number of images or blocks which can be packed in each ciphertext.

\paragraph{\bf{Trade-offs for the entropy of the secret key}}
There exists a trade-off in terms of the secret key generation for resisting distinguishing attacks and birthday attacks \cite{PTP2015}: the secret key must have a large enough variance to resist a birthday attack (more key entropy), but small enough to be resilient against a distinguishing attack (lower root Hermite factor). That is, if we consider that all the parameters are constant except for the variance of the secret key, we have asymptotically $\frac{q}{s} \propto \frac{s^{2D + 2}}{s} = s^{2D+1}$ (see Eq.~\eqref{eq:delta}). So if we increase $s$, the entropy of the secret key increases along with $\delta$ (the converse is also true: reducing the variance of $s$ reduces the entropy and $\delta$). Nevertheless, the size of the key $n=\prod_{i=0}^{m-1}n_i$ has a higher impact on the root Hermite factor $\delta$ than the deviation per component ($\sigma=s/\sqrt{2\pi}$) (cf. Eq.~\eqref{eq:delta}), and that is the reason why increasing $m$ or $n_i$ in $m$-RLWE provides a higher security against distinguishing attacks than the RLWE counterpart with the same key entropy, as the used polynomials have more coefficients.\footnote{A factor of $\mathcal{O}(I)$ more coefficients with respect to $2$-RLWE, and $\mathcal{O}(IN)$ coefficients with respect to an RLWE solution}

If the key entropy from RLWE is already enough against birthday attacks, there is no need to increase it further when transitioning to $m$-RLWE, so it is possible to reduce the variance of the Gaussian variables that generate the $m$-RLWE secret key. This can be optimized for obtaining a secret key with the same entropy as for the previous cryptosystems~\cite{PTP2015} and \cite{LNV11} while, at the same time, still reducing the value of $\delta$ (thus, improving the security against distinguishing attacks). 

\subsection{Computational cost for the new encryption and decryption primitives}

In the previous sections we compared the computational cost for encrypted linear convolutions (correlation and filtering), for which such cost is proportional to a constant number of products between ring elements belonging to $R_q$. These results are applicable when all blocks are processed identically (See Section~\ref{sec:blockprocess}), as encryption and decryption are unaffected. When the process is different for each block, we have proposed a modification for encryption and decryption by introducing pre- and post-processing in them (see Section~\ref{sec:prepostprocess}).
We now analyze the impact of such pre- and post-processing in terms of computational cost. Returning to the example of filtering between $I$ images with size $N \times N$ and filters of size $F \times F$, the cost for the product between polynomials from our cryptosystem is
\begin{equation*}
\mbox{Cost}_{Poly.Prod} \approx {(N + F - 1)}^4h^2I^2.
\end{equation*}

On the other hand, the cost of a pre- or post-processing operation would be $(N^2 + F^2)\mbox{Cost}_{DFT(I points)}$ because we have to perform $N^2$ DFTs of size $I$ for the images and $F^2$ DFTs of size $I$ for the filters. If we use a fast algorithm like the FFT for computing the polynomial products and the DFT, we will have a total cost of
\begin{equation*}
\begin{split}
\mbox{Cost} & \approx \\
 & N_{Poly. Prod.}C_{FFT}{(N + F - 1)}^2hI \log_2{((N + F - 1)^2hI)} \\
 & + (N^2 + F^2)C_{FFT}I \log_2I,
\end{split}
\end{equation*}
where $N_{Poly. Prod.}$ is the number of polynomial products needed for performing the considered cryptographic primitive (in this case, encryption or decryption), and $C_{FFT}$ is the linear constant of the used FFT algorithm.

Using a slack value of $h = 1$, we can obtain the ratio between the cost for the pre- or post-processing and the respective encryption/decryption primitive (with no pre-/post-processing):
\begin{equation*}
\begin{split}
 & \mbox{Ratio}_{Cost} \approx \\
 & \frac{(N^2 + F^2) C_{FFT}I \log_2 I}{N_{Poly. Prod.}C_{FFT}{(N + F - 1)}^2 I \log_2 ({(N + F - 1)}^2I)},
\end{split}
\end{equation*}
where $\mbox{Ratio}_{Cost}$ achieves its highest value when $F = 1$.

Now, let us express the asymptotic $\mbox{Ratio}_{Cost}$ when $F = 1$ and $N \rightarrow \infty$:
\begin{equation*}
\begin{split}
\lim_{N \to \infty} \mbox{Ratio}_{Cost} = & \lim_{N \to \infty} \frac{(N^2 + 1) C_{FFT}I \log_2 I}{N_{Poly. Prod.}C_{FFT}{N}^2 I \log_2 ({N}^2I)} \\
       = &\lim_{N \to \infty} \frac{(1 + \frac{1}{N^2}) C_{FFT}I \log_2 I}{N_{Poly. Prod.}C_{FFT} I \log_2 ({N}^2I)} \\
       = & 0. 
\end{split}
\end{equation*}

Therefore, when increasing the size of the images, the additional cost for the primitives becomes negligible. Additionally, it is also interesting to calculate the maximum increase in computational cost that the use of pre- and post-processing can incur on. With this aim, we study the case when $I \rightarrow \infty$ and $F = 1$:
\begin{equation*}
\begin{split}
\lim_{I \to \infty} \mbox{Ratio}_{Cost} = & \\
       = &\lim_{I \to \infty} \frac{(N^2 + 1) C_{FFT}I \log_2 I}{N_{Poly. Prod.}C_{FFT}{N}^2 I \log_2 ({N}^2I)} \\
                                 = & \lim_{I \to \infty} \frac{N^2 + 1}{N_{Poly. Prod.}N^2\frac{\log_2{IN^2}}{\log_2I}} \\
                                 = & \lim_{I \to \infty} \frac{N^2 + 1}{N_{Poly. Prod.}N^2 (\frac{\log_2{N^2}}{\log_2I} + 1)} \\
                                 = & \frac{N^2 + 1}{N_{Poly. Prod.}N^2},
\end{split}                                 
\end{equation*}
that is approximately $\frac{1}{N_{Poly. Prod.}}$ when $N$ is big enough.

Hence, the worst-case computational cost of the modified encryption and decryption primitives with respect to the original one is $\mbox{Cost}_{Orig.Primitive} \cdot (1 + \frac{1}{N_{Poly. Prod.}})$. The encryption conveys $2$ polynomial products, so $\frac{3}{2}\mbox{Cost}_{Encryption}$ ($\mbox{Cost}_{Encryption}$ represents the computational cost of the original encryption), and for the decryption it depends on both the number of polynomial elements comprising the ciphertexts and the computation of the powers of the secret key. Assuming that the powers of the secret key have been precomputed, we would have $\mbox{Cost}_{Decryption} \cdot (1 + \frac{1}{\mbox{Num. of Elements} - 1}) = \mbox{Cost}_{Decryption}\frac{\mbox{Num. of Elements}}{\mbox{Num. of Elements} - 1}$ ($\mbox{Cost}_{Decryption}$ represents the computational cost of the original decryption).

Summarizing, we can see that the cost increase due to the use of the pre and post-processing is very small, and in fact, it becomes negligible for practical cases.

\subsection{Implementation and execution times}

We have implemented both Lauter RLWE-based cryptosystem and our $m$-RLWE extension in C using the GMP 6.0.0\footnote{``GNU Multiple Precision Arithmetic Library,'' \url{www.gmplib.org}.} and NFLlib~\cite{ABGGKL16} libraries. Table~\ref{tab:tableexp1} compares the obtained encrypted filtering runtimes with a square filter of side $F=11$ on an Intel Xeon E5-2620 processor running Linux; we consider several ($I$) packed images of size $N \times N$ in the same ciphertext with: a) the traditional Paillier (with a clear text filter), b) a RLWE cryptosystem~\cite{LNV11}, c) a 2-RLWE cryptosystem~\cite{PTP2015}, and d) its $3$-RLWE counterpart.

The reported encryption times comprise the encryption of all involved signals, both images and filters, except for Paillier, for which the filters are not encrypted. We do not include relinearization steps after each multiplication, and instead take into account the more demanding decryption of the extended encryptions (ciphertexts grow after each multiplication if relinearization is not applied).

\begin{table}[!t]
\renewcommand{\arraystretch}{1.3}
\caption{Encrypted filtering performance ($D=1$, $t=12289$, $s=\sqrt{2\pi}$)}
\label{tab:tableexp1}
\centering \ssmall
\begin{tabular}{|c| c c c|}\hline
$(I, N)$                             & (4, 246)      & (2, 502)       & (4, 502) \\
\hline
\hline
\multicolumn{4}{|c|}{$3$-RLWE cryptosystem}\\
\hline
 $n$                          & 262144    & 524288    & 1048576 \\
 $\lceil \log_2 (q) \rceil$        & 59       & 61        & 62 \\
 Enc. images size (bits)  & 6.15$\cdot$10$^7$     & 1.26$\cdot$10$^8$     & 2.59$\cdot$10$^8$ \\
 $\delta$                     & 1.000039  & 1.000020  & 1.000010 \\
 Bit security                 & 31968  & 62451  & 121978 \\
 Encrypt. time (\textit{ms})            & 67 & 137 & 300 \\
 Decrypt. time (\textit{ms})            & 16 & 33 & 87 \\
 Conv. time (\textit{ms})            & 7 & 14 & 29 \\
\hline
\hline 
\multicolumn{4}{|c|}{$2$-RLWE cryptosystem}\\
\hline
 $n$                           & 65536    & 262144    & 262144 \\
 $\lceil \log_2 (q) \rceil$       & 56       & 59        & 59 \\
 Enc. images size (bits)  & 5.84$\cdot$10$^7$     & 1.23$\cdot$10$^8$     & 2.46$\cdot$10$^8$ \\
 $\delta$                     & 1.00015  & 1.000039  & 1.000039 \\
 Bit security                 & 8340  & 31968  & 31968 \\
 Encrypt. time (\textit{ms})            & 61 & 128 & 257 \\
 Decrypt. time (\textit{ms})            & 13 & 29 & 57 \\
 Conv. time (\textit{ms})            & 6 & 14 & 27 \\
\hline
\hline
\multicolumn{4}{|c|}{RLWE cryptosystem}\\
\hline 
 
 $n$                            & 2048 (h = 8)    & 2048 (h = 4)    & 2048 (h = 4) \\
 $\lceil \log_2 (q) \rceil$       & 49      & 49      & 49 \\
 Enc. images size (bits)  & 2.03$\cdot$10$^8$     & 2.02$\cdot$10$^8$     & 4.05$\cdot$10$^8$ \\
 $\delta$                   & 1.0041  & 1.0041  & 1.0041 \\
 Bit security               & 195  & 195  & 195 \\
 Encrypt. time (\textit{ms})            & 236 & 235 & 470 \\
 Decrypt. time (\textit{ms})            & 81 & 81 & 162 \\
 Conv. time (\textit{ms})           & 639 & 652 & 1305 \\

\hline
\hline
\multicolumn{4}{|c|}{Paillier cryptosystem (with 2048 bit modulus, 112 bits of security)}\\
\hline 
 Enc. images size (\textit{bits})           & $9.92\cdot10^8$ & $2.06\cdot10^9$ & $4.13\cdot10^9$ \\
 Encrypt. time (\textit{s})            & $3.36\cdot10^3$ & $6.99\cdot10^3$ & $13.98\cdot10^3$ \\
 Decrypt. time (\textit{s})            & $3.61\cdot10^3$ & $7.23\cdot10^3$ & $14.46\cdot10^3$ \\
 Conv. time (\textit{s})           & $2.16\cdot10^3$ & $4.51\cdot10^3$ & $9.01\cdot10^3$ \\
\hline
\end{tabular}
\end{table}

Table~\ref{tab:tableexp2} compares the encrypted image filtering performance considering both 3D volumetric images (as the ones used in MRI applications) of size $N_x \times N_y \times N_z$ (with $N_z = 12$ for all the considered images), and a 3D Gaussian smoothing kernel (which is not encrypted) of length $F=5$ in each dimension with: a) traditional Paillier, b) a RLWE cryptosystem~\cite{LNV11}, c) a 2-RLWE cryptosystem~\cite{PTP2015}, and d) its $3$-RLWE counterpart.

\begin{table}[!t]
\renewcommand{\arraystretch}{1.3}
\caption{Encrypted 3D Gaussian Smoothing ($D=1$, $t=12289$, $s=\sqrt{2\pi}$)}
\label{tab:tableexp2}
\centering \ssmall
\begin{tabular}{|c| c c c|}\hline
$N_x \times N_y$  & $60 \times 60$  & $124 \times 124$       & $252 \times 252$ \\
\hline
\hline
\multicolumn{4}{|c|}{$3$-RLWE cryptosystem}\\
\hline
 $n$                           & 65536    & 262144    & 1048576 \\
 $\lceil \log_2 (q) \rceil$       & 56       & 59        & 62 \\
 Enc. image size (bits) & 7.30$\cdot$10$^6$     & 3.08$\cdot$10$^7$     & 1.29$\cdot$10$^8$ \\
 $\delta$                     & 1.00015  & 1.000039  & 1.000010 \\
 Bit security                 & 8340  & 31968  & 121978 \\
 Encrypt. time (\textit{ms})            & 8 & 32 & 146 \\
 Decrypt. time (\textit{ms})            & 2 & 11 & 69 \\
 Conv. time (\textit{ms})            & 1 & 4 & 16 \\
\hline
\hline 
\multicolumn{4}{|c|}{$2$-RLWE cryptosystem}\\
\hline
 $n$                            & 4096    & 16384    & 65536 \\
 $\lceil \log_2 (q) \rceil$        & 50       & 53        & 56 \\
 Enc. image size (bits)  & 4.88$\cdot$10$^6$     & 2.07$\cdot$10$^7$     & 8.76$\cdot$10$^7$ \\
 $\delta$                     & 1.0021  & 1.00056  & 1.00015 \\
 Bit security                 & 481  & 2122  & 8340 \\
 Encrypt. time (\textit{ms})            & 5 & 22 & 91 \\
 Decrypt. time (\textit{ms})            & 2 & 8 & 39 \\
 Conv. time (\textit{ms})           & 4 & 16 & 79 \\
\hline
\hline
\multicolumn{4}{|c|}{RLWE cryptosystem}\\
\hline 
 
 $n$                            & 2048 ($h = 32$)    & 2048 ($h = 16$)    & 2048 ($h = 8$) \\
 $\lceil \log_2 (q) \rceil$       & 49      & 49      & 49 \\
 Enc. image size (bits)  & 1.42$\cdot$10$^8$     & 2.94$\cdot$10$^8$     & 5.97$\cdot$10$^8$ \\
 $\delta$                     & 1.0041  & 1.0041  & 1.0041 \\
 Bit security                 & 195  & 195  & 195 \\
 Encrypt. time (\textit{ms})            & 165 & 342 & 695 \\
 Decrypt. time (\textit{ms})            & 57 & 115 & 229 \\
 Conv. time (\textit{ms})           & 587 & 1213 & 2466 \\

\hline
\hline
\multicolumn{4}{|c|}{Paillier cryptosystem (with 2048 bit modulus, 112 bits of security)}\\
\hline 
 Enc. image size (\textit{bits})           & $1.77\cdot10^8$ & $7.56\cdot10^8$ & $3.12\cdot10^9$ \\
 Encrypt. time (\textit{s})            & 599 & 2560 & 10572 \\
 Decrypt. time (\textit{s})            & 904 & 3614 & 14458 \\
 Conv. time (\textit{s})          & 399 & 1704 & 7037 \\
\hline
\end{tabular}
\end{table}
Tables~\ref{tab:tableexp1} and~\ref{tab:tableexp2} show that the runtimes of lattice-based cryptosystems clearly outperform those provided by Paillier; both $3$-RLWE and $2$-RLWE schemes are the fastest and the most compact in terms of cipher expansion. Even though the $2$-RLWE cryptosystem is slightly faster and has less cipher expansion than the $3$-RLWE counterpart, the security provided by the latter is much higher, so incorporating ``virtual'' dimensions contributes to increasing the security without significantly impacting efficiency.

Table~\ref{tab:tableexp3} compares the homomorphic computation of the $8 \times 8$ block-DCT (the DCT transform matrix is known and, hence, public) for images of size $N \times N$, with: a) traditional Paillier, b) a RLWE cryptosystem~\cite{LNV11}, and c) a 2-RLWE cryptosystem. The  corresponding sizes of the relinearization matrices are also reported.
Table~\ref{tab:tableexp3} shows that even though all lattice-based schemes outperform the use of Paillier for encryption/decryption, the block DCT transforms cannot be efficiently computed with an RLWE-based solution due to the huge size of the relinearization matrix. By considering a ``virtual'' dimension with $2$-RLWE, this size is reduced in several orders of magnitude, considerably improving the runtimes for the block DCT transforms.

\begin{table}[!t]
\renewcommand{\arraystretch}{1.3}
\caption{Encrypted Block-DCT performance ($D=1$, $t=12289$, $s=\sqrt{2\pi}$)}
\label{tab:tableexp3}
\centering \ssmall
\begin{tabular}{|c| c c c c|}\hline
$N$                        & 64      & 128      & 256       & 512 \\
\hline 
\hline
\multicolumn{5}{|c|}{$2$-RLWE cryptosystem}\\
\hline
 $n$                        & 4096    & 16384    & 65536    & 262144 \\
 $\lceil \log_2 (q) \rceil$ & 50       & 53       & 56        & 59 \\
 Enc. image size (bits) & 4.07$\cdot$10$^5$ & 1.73$\cdot$10$^6$     & 7.30$\cdot$10$^6$     & 3.08$\cdot$10$^7$ \\
 Matrix Relin. (bits) & 1.05$\cdot$10$^8$  & 4.45$\cdot$10$^8$  & 2.35$\cdot$10$^9$  & 9.90$\cdot$10$^9$ \\
 $\delta$                   & 1.0021  & 1.00056  & 1.00015  & 1.000039 \\
 Bit security               & 481  & 2122  & 8340  & 31968 \\
 Encrypt. time (\textit{ms})           & 0.46 & 1.81 & 7.61 & 32.10 \\
 Decrypt. time (\textit{ms})           & 0.12 & 0.52 & 2.43 & 11.19 \\
 Block-DCT (\textit{s})          & 0.16  & 0.66 & 2.87 & 11.83 \\
\hline
\hline
\multicolumn{5}{|c|}{RLWE cryptosystem}\\
\hline 
 
 $n$                        & 4096    & 16384    & 65536    & 262144 \\
 $\lceil \log_2 (q) \rceil$ & 50      & 53      & 56      & 59 \\
 Enc. image size (bits) & 4.07$\cdot$10$^5$ & 1.73$\cdot$10$^6$     & 7.30$\cdot$10$^6$     & 3.08$\cdot$10$^7$ \\
 Matrix Relin. (bits) & 6.71$\cdot$10$^9$  & 1.14$\cdot$10$^{11}$  & 2.41$\cdot$10$^{12}$  & 4.05$\cdot$10$^{13}$ \\
 $\delta$                   & 1.0021  & 1.00056  & 1.00015  & 1.000039 \\
 Bit security               & 481  & 2122  & 8340  & 31968 \\
 Encrypt. time (\textit{ms})           & 0.46 & 1.81 & 7.61 & 32.10 \\
 Decrypt. time (\textit{ms})           & 0.12 & 0.52 & 2.43 & 11.19 \\
 Block-DCT time (\textit{s})          & 10.28 & 167.82 & 2940.47 & 48440.75 \\

\hline
\hline
\multicolumn{5}{|c|}{Paillier cryptosystem (with 2048 bit modulus, 112 bits of security)}\\
\hline 
 Enc. image size (\textit{bits})          & $1.68\cdot10^7$ & $6.71\cdot10^7$ & $2.68\cdot10^8$ & $1.07\cdot10^9$ \\
 Encrypt. time (\textit{s})           & 57 & 227 & 909 & 3637 \\
 Decrypt. time (\textit{s})           & 56 & 226 & 904 & 3614 \\
 Block-DCT time (\textit{s})          & 4.79 & 19.15 & 76.59 & 306.37 \\
\hline
\end{tabular}
\end{table}

\section{Conclusions}
\label{sec:conclusions}
We have presented a novel hard problem, denoted Multivariate Ring Learning with Errors ($m$-RLWE), that enables efficient encrypted processing of images and multidimensional signals (3-D images, video,...). Cryptosystems based on this problem can flexibly fit the input signal structure, therefore producing an extremely efficient encryption with very low processing overhead and cipher expansion. We have also produced novel techniques to deal with non-interactive transformations between different structures, enabling for the first time block-based multidimensional encrypted signal processing in a non-interactive way. This is especially relevant in privacy-aware scenarios like outsourced medical imaging (ECG, EEG, CT scans, MRI,...), where the proposed solutions enable unprecedented encrypted performance and security levels. The proposed problem, encryption mechanisms and transformation techniques open up a wide range of novel encrypted processing applications supporting secure unattended outsourced processing of signals of almost any kind. 

\ifCLASSOPTIONcaptionsoff
  \newpage
\fi

\bibliographystyle{IEEEtran}
\bibliography{bibliografia}

\end{document}